\documentclass[pdflatex,sn-mathphys]{sn-jnl}

\jyear{2021}%
\usepackage{siunitx}

\theoremstyle{thmstyleone}%
\theoremstyle{thmstyletwo}%

\theoremstyle{thmstylethree}%

\begin{document}

\title[Proton Irradiation of SiPM arrays for POLAR-2]{Proton Irradiation of SiPM arrays for POLAR-2}


\author*[1]{\fnm{Slawomir} \sur{Mianowski}}\email{slawomir.mianowski@ncbj.gov.pl}
\author*[2]{\fnm{Nicolas} \sur{De Angelis}}\email{nicolas.deangelis@unige.ch}
\author*[2]{\fnm{Johannes} \sur{Hulsman}}\email{johannes.hulsman@unige.ch}

\author[2]{\fnm{Merlin} \sur{Kole}}\email{merlin.kole@unige.ch}
\author[3]{\fnm{Tomasz} \sur{Kowalski}}\email{tomasz.kowalski@ifj.edu.pl}
\author[3]{\fnm{Sebastian} \sur{Kusyk}}\email{sebastian.kusyk@ifj.edu.pl}
\author[4]{\fnm{Hancheng} \sur{Li}}\email{hancheng.li@unige.ch}
\author[1]{\fnm{Zuzanna} \sur{Mianowska}}\email{zuzanna.mianowska@ncbj.gov.pl}
\author[3]{\fnm{Jerzy} \sur{Mietelski}}\email{jerzy.mietelski@ifj.edu.pl}
\author[1,6]{\fnm{Agnieszka} \sur{Pollo}}\email{agnieszka.pollo@ncbj.gov.pl}
\author[1]{\fnm{Dominik} \sur{Rybka}}\email{dominik.rybka@ncbj.gov.pl}
\author[5]{\fnm{Jianchao} \sur{Sun}}\email{sunjc@ihep.ac.cn}
\author[3]{\fnm{Jan} \sur{Swakon}}\email{jan.swakon@ifj.edu.pl}
\author[3]{\fnm{Damian} \sur{Wrobel}}\email{damian.wrobel@ifj.edu.pl}
\author[2]{\fnm{Xin} \sur{Wu}}\email{xin.wu@unige.ch}

\affil*[1]{\orgname{National Centre for Nuclear Research}, \orgaddress{\street{A. Soltana 7 Street}, \city{Otwock}, \postcode{PL-05400}, \country{Poland}}}

\affil*[2]{\orgdiv{DPNC}, \orgname{University of Geneva}, \orgaddress{\street{24 Quai Ernest-Ansermet}, \city{Geneva}, \postcode{CH-1205}, \country{Switzerland}}}

\affil[3]{\orgdiv{Institute of Nuclear Physics}, \orgname{Polish Academy of Sciences}, \orgaddress{\street{Radzikowskiego 152 Street}, \city{Krakow}, \postcode{PL-31342}, \country{Poland}}}

\affil[4]{\orgdiv{Geneva Observatory, ISDC}, \orgname{University of Geneva}, \orgaddress{\street{16 Chemin d’Ecogia}, \city{Versoix}, \postcode{CH-1290}, \country{Switzerland}}}

\affil[5]{\orgdiv{Institute of High Energy Physics}, \orgname{Chinese Academy of Sciences CN}, \orgaddress{\street{19B Yuquan Road}, \city{Beijing}, \postcode{100049}, \country{China}}}

\affil[6]{\orgname{Astronomical Observatory of the Jagiellonian University}, \orgaddress{\street{Orla 171 Street}, \city{Krakow}, \postcode{PL-30244}, \country{Poland}}}

\abstract{POLAR-2 is a space-borne polarimeter, built to investigate the polarization of Gamma-Ray Bursts and help elucidate their mechanisms. The instrument is targeted for launch in 2024 or 2025 aboard the China Space Station and is being developed by a collaboration between institutes from Switzerland, Germany, Poland and China.

The instrument will orbit at altitudes between \SI{340}{km} and \SI{450}{km} with an inclination of \SI{42}{^{\circ}} and will be subjected to background radiation from cosmic rays and solar events. It is therefore pertinent to better understand the performance of sensitive devices under space-like conditions.

In this paper we focus on the radiation damage of the silicon photomultiplier arrays S13361-6075NE-04 and S14161-6050HS-04 from Hamamatsu. The S13361 are irradiated with \SI{58}{MeV} protons at several doses up to \SI{4.96}{Gy}, whereas the newer series S14161 are irradiated at doses of \SI{0.254}{Gy} and \SI{2.31}{Gy}. Their respective performance degradation due to radiation damage are discussed. The equivalent exposure time in space for silicon photomultipliers inside POLAR-2 with a dose of \SI{4.96}{Gy} is \SI{62.9}{years} (or \SI{1.78}{years} when disregarding the shielding from the instrument).
Primary characteristics of the I-V curves are an increase in the dark current and dark counts, mostly through cross-talk events. 
Annealing processes at $\SI{25}{^{\circ}C}$ were observed but not studied in further detail. Biasing channels while being irradiated have not resulted in any significant impact. 

Activation analyses showed a dominant contribution of $\beta^{+}$ particles around \SI{511}{keV}. These resulted primarily from copper and carbon, mostly with decay times shorter than the orbital period. \newline}




\maketitle

\section{Motivation} \label{sec:motiv}

Gamma-Ray Bursts (GRBs) are the most energetic astrophysical events in the Universe known to man, emitting energies up to $\SI{E53}{erg}$. They have a short prompt emission which is followed by a longer lived afterglow with wavelengths ranging from radio waves to TeV energies. Mergers of compact binaries were long thought to generate short GRBs, which was experimentally confirmed by the joint detection of GRB170817A by LIGO, VIRGO, INTEGRAL and \textit{Fermi}-GBM \cite{Abbott_2017}. Regarding long GRBs, growing evidence indicates that they are accompanied by supernovae events \cite{Woosley_2006}.

Benefits of gamma-ray polarization studies have spawned numerous polarimetry missions \cite{Yonetoku_2011,Produit_2018,Lowell_2017,Chattopadhyay_2019}, the most dedicated of which are GAP (50-$\SI{300}{keV}$) \cite{Yonetoku_2011} and POLAR (50-$\SI{500}{keV}$) \cite{Zhang_2019}, which is the predecessor to POLAR-2 \cite{POLAR-2:2021uea}. Extensive polarization analyses by POLAR are discussed in their catalogue paper \cite{kole2020polar}. Despite the scientific advances in GRB polarimetry, no analysis can compensate for the need of higher precision measurements. 
A mission with improved instrumentation and longer lifetime, amongst which finer time and/or energy resolved polarization analyses can be carried out, will obtain larger GRB samples. POLAR-2 is dedicated to that cause, accounting for the experiences gained from its predecessor. It will be launched towards the China Space Station (CSS) in 2024 or 2025 for a mission of at least 2 years. The CSS (and hence POLAR-2) is orbiting at a typical altitude of \SI{383}{km} at an inclination of 42$^{\circ}$. As a result, it is exposed to radiation from cosmic rays of galactic, solar and trapped origin (more to it in section \ref{sec:leo_bkg}). 

Silicon photomultipliers (SiPMs) are chosen as photodetectors for POLAR-2. They are photon counting devices wherein multiple avalanche photodiode pixels operating in Geiger mode. Generally, SiPMs provide a high gain (10$^5$ to 10$^7$) comparable with classical photomultiplier tubes, have a fast response time, are compact, require a low bias voltage (below \SI{100}{V}) and are insensitive to a magnetic field. However, when exposed to radiation, they can suffer from either bulk damage (primarily due to non-ionizing energy loss - NIEL) or surface damage (primarily due to ionizing energy loss - IEL) \cite{GARUTTI201969}, resulting in a decreased scientific performance of the instrument. These effects are minimized (yet not prevented) through shielding of the mechanical structure of the instrument. Regardless, dedicated measurements are necessary to estimate the long term performance of such components like SiPM arrays and how this affects the science case of POLAR-2. These will be discussed in this paper.\newline 

\newpage

\section{Instrument Design} \label{sec:instrument_design}

\begin{figure}[!h]
     \centering
     \begin{subfigure}[b]{0.32\textwidth}
         \centering
         \includegraphics[width=\textwidth]{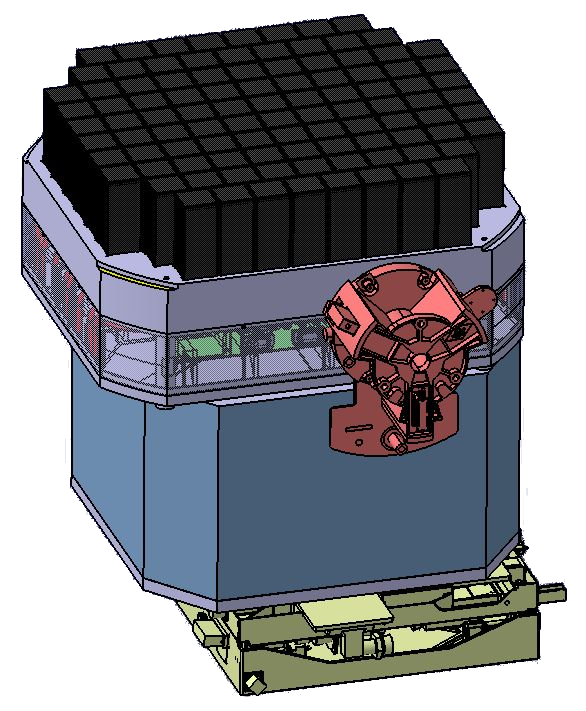}
         \caption{}
         \label{fig:polar2_cad}
     \end{subfigure}
     \begin{subfigure}[b]{0.32\textwidth}
         \centering
         \includegraphics[width=0.4\textwidth,angle=180]{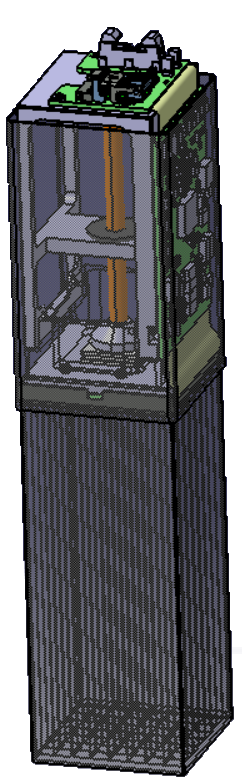}
         \caption{}
         \label{fig:module_vertical}
     \end{subfigure}
     \begin{subfigure}[b]{0.32\textwidth}
         \centering
         \includegraphics[width=\textwidth]{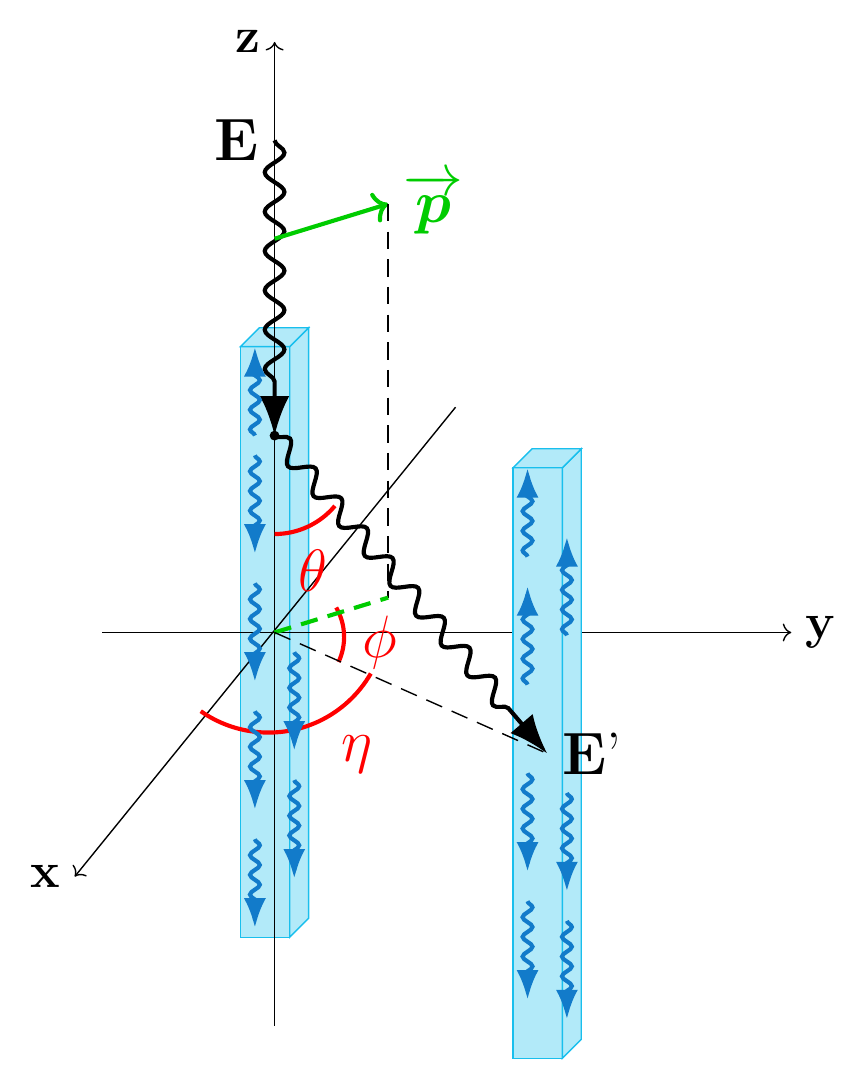}
         \caption{}
         \label{fig:compton_scatter}
     \end{subfigure}
        \caption{\textbf{a)} A preliminary CAD model of POLAR-2 and \textbf{b)} the polarimeter module \cite{POLAR-2:2021uea}. \textbf{c)} A schematic of a photon scattering between two scintillator bars. These bars do not necessarily need to be inside the same module.}
        \label{fig:POLAR2_design}
\end{figure}

Polarization measurements of $\gamma$-rays can be achieved through Compton scattering, whose differential cross section is described by the Klein-Nishina formula \cite{1929ZPhy...52..853K}. Its azimuthal scattering angle is linked to the least populated scattering angle in the distribution (referred to as the polarization angle) as the photon preferentially scatters orthogonal to the polarization vector. This phenomenon can be exploited by developing an instrument with segmented (low-Z) scintillator bars, where a $\gamma$ photon scatters in the first bar and is absorbed/scattered by a second; as illustrated in Figure~\ref{fig:compton_scatter}. The relative position between these bars allows one to derive the scattering angle. Numerous photons from a GRB will lead to a scattering angle distribution, also referred to as a modulation curve, whose shape is used to determine the polarization angle, which is related to the phase of a $180^\circ$ modulation, and the polarization degree which is related to the amplitude of this modulation. \newline

\begin{figure}[!h]
  \centering
  \includegraphics[scale=0.4]{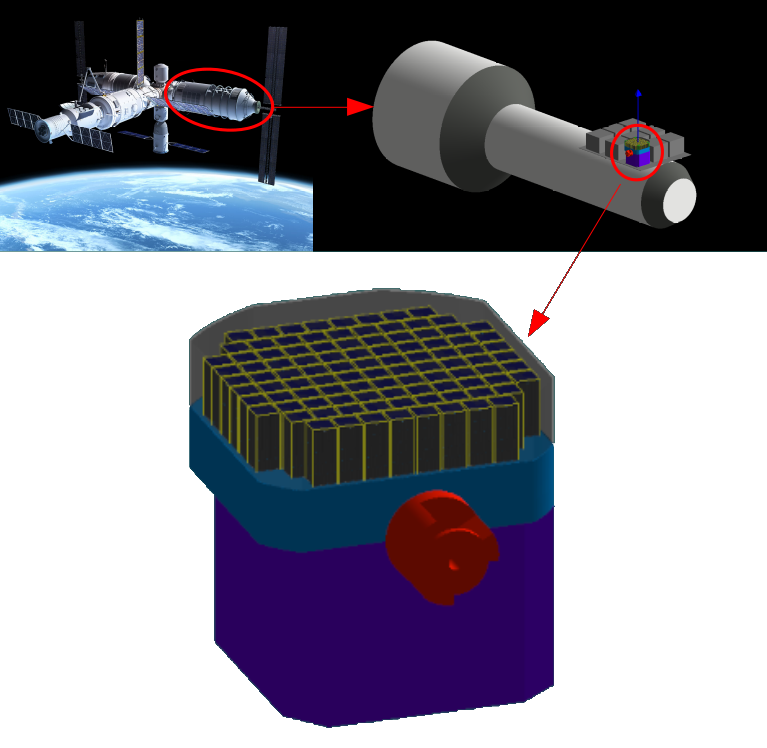}
  \caption{\textbf{Top left} shows a 3D rendering of the CSS. \textbf{Top right} provides a simplified version of the experimental module, whereas in the \textbf{bottom} you can see the full POLAR-2 instrument in GEANT4 \cite{AGOSTINELLI2003250}. }
  \label{fig:polar2_g4}
\end{figure}

POLAR-2, the successor to POLAR \cite{Produit_2018}, applies the aforementioned technique into its polarimetry instrument. The full instrument, as shown in Figure~\ref{fig:polar2_cad}, encompasses a total volume of about \SI{60}{cm}$\times$\SI{60}{cm}$\times$\SI{70}{cm}. The polarimeter consists of 100 modules facing deep space with a half-sky field of view. Each detector module, seen in Figure~\ref{fig:module_vertical}, consists of a target of 64 plastic scintillators with dimensions \SI{5.9}{mm}$\times$\SI{5.9}{mm}$\times$\SI{125}{mm}. Each scintillator is read out on the bottom side by its own SiPM channel. The 64 SiPMs in turn are readout using the module front-end electronics which handles the trigger logic and communication of the detector module with the back-end electronics. The top of the scintillators is covered by a sorbothane damper system and the full module is covered in a carbon fibre socket. Additionally, all modules together are covered by one large carbon fibre shield as shown in Figure~\ref{fig:polar2_g4}\footnote{The amount of shielding in the zenith direction is therefore \SI{125}{mm} of plastic scintillator, \SI{1}{mm} of sorbothane dampers, and \SI{4}{mm} of carbon fiber housing.}. Below the detector modules POLAR-2 consists of a large aluminium structure which houses the back-end and power supplies as well as the communication system with the CSS. 

Contrary to its predecessor POLAR, where the scintillators were read-out using a multi-anode Photomultiplier Tubes (MAPMT), the 6400 plastic scintillators are read out by 400 16-channel SiPM arrays, also called Muli-Pixel Photon Counter (MPPC), from Hamamatsu Photonics (S13361-6075NE-04, referred to as S13361 in this paper). 
This change serves three main purposes. First, it allows for a more robust mechanical design, as unlike PMTs, SiPMs do not have a glass window. Second, using SiPMs  (instead of PMTs) removes the need for a high voltage power supply which is complex to design and operate in space. Finally, the SiPM readout increases the sensitivity. By using SiPMs the number of photo-electrons (p.e.) per deposited keV in the scintillators can be increased from 0.3 p.e./keV (in the case of POLAR) to 1.6 p.e./keV. As a result, we achieve a lower energy detection threshold. \newline

The typical downside of SiPM use over standard PMTs is the dark noise. The dark noise rate is proportional to the SiPM temperature. As such, great care was taken in the POLAR-2 design to minimize the temperature of the SiPM arrays. Based on current thermal simulations the mean temperature, as measured on in the centre of each SiPM array, is predicted to be $-10^\circ$C. This is achieved by both providing a good thermal contact to the radiators of the POLAR-2 instrument, through copper pipes, as well as by using a Peltier element directly connected to the SiPM. The Peltier element not only cools the SiPM array by several degrees, it also allows to stabilize the temperature of the SiPMs, during an orbit, to within several degrees through an active control system. This system is based on a PT100 sensor placed on the SiPMs, readout by an FPGA on the front-end electronics which in turn controls the current to the Peltier element. In addition, the bias voltage of the SiPM is controlled by the FPGA in a similar fashion using the input from the PT100 sensor.This allows for a constant gain the SiPM during the mission.

Apart from the temperature the dark noise also increases significantly for high radiation dose. This effect must be studied carefully for better low threshold energy determination. Additionally, the main challenge in polarization measurements is to achieve a high level of detection uniformity, including SiPM single channel and scintillator response. For POLAR-2 this implies a uniform response over the 6400 detector channels in order to allow to distinguish between the non-uniformity induced by a potential polarization signal and that induced by a non-uniform detector response. The scheme of the proposed polarimeter shows that one would expect the outside of the instrument to suffer more from cosmic radiation than the centre, the induced damage will therefore not be uniform, resulting potentially in a higher noise level for channels on the outside of the instrument compared to those in the centre. It is therefore important to understand in detail the deterioration of SiPMs performance in order to allow to mitigate when possible or model noise effect in detector simulations. \newline

A dedicated POLAR-2 Monte Carlo simulation framework was developed for detector science analyses. Here, the GEANT4 framework \cite{AGOSTINELLI2003250}, v4.10.07.p02, is used to characterize the instrument as well as its response\footnote{This will be outlined in more details in a future publication.}. The following physics lists have been used for subsequent background radiation simulations: \textit{G4EmLivermorePolarizedPhysics}, \textit{G4EmExtraPhysics}, \textit{G4DecayPhysics}, \textit{G4RadioactiveDecayPhysics}, \textit{G4HadronElasticPhysicsHP}, \textit{G4StoppingPhysics}, \textit{G4IonQMDPhysics}, \textit{G4IonElasticPhysics}, \textit{G4HadronPhysicsQGSP\_BERT\_HP}, \textit{GammaNuclearPhysics} and \textit{G4EmStandardPhysics}. \newline

The design illustrated in Figure~\ref{fig:polar2_g4} reflects the most up-to-date design at the time of this writing. A 3D rendering of the CSS can also be seen there. As encircled, POLAR-2 is situated on one of the experimental modules of the space station (on the \textit{Wentian} module which was launched on July 24th 2022). A simplified version of the experimental module was characterized in GEANT4 to account for the shadow cast by it\footnote{A more detailed version will be presented in the future when the CSS' design is completed.}. Its wall thickness is \SI{10}{mm} and is composed of aluminium (alloy 2219). 8 neighbouring instrument are also included. Their design is unknown and are therefore represented by $600\times600\times$\SI{500}{mm^3} solid aluminium (same alloy as the space station) boxes. At the bottom of the figure the implementation of the POLAR-2 instrument in Geant4 can be recognized. It is nearly identical to that in Figure~\ref{fig:polar2_cad} (with the exception of the missing carbon fiber shielding). For clarity, each single SiPM channel is modelled by a $\SI{100}{\mu m}$ epoxy resin and $\SI{450}{\mu m}$ of silicon \footnote{Exact specification was not delivered by Hamamatsu.}.

\section{LEO Background \& Anticipated Radiation Doses} \label{sec:leo_bkg}

\begin{figure}[t]
  \centering
  \begin{subfigure}[b]{0.49\textwidth}
     \centering
     \includegraphics[width=\textwidth]{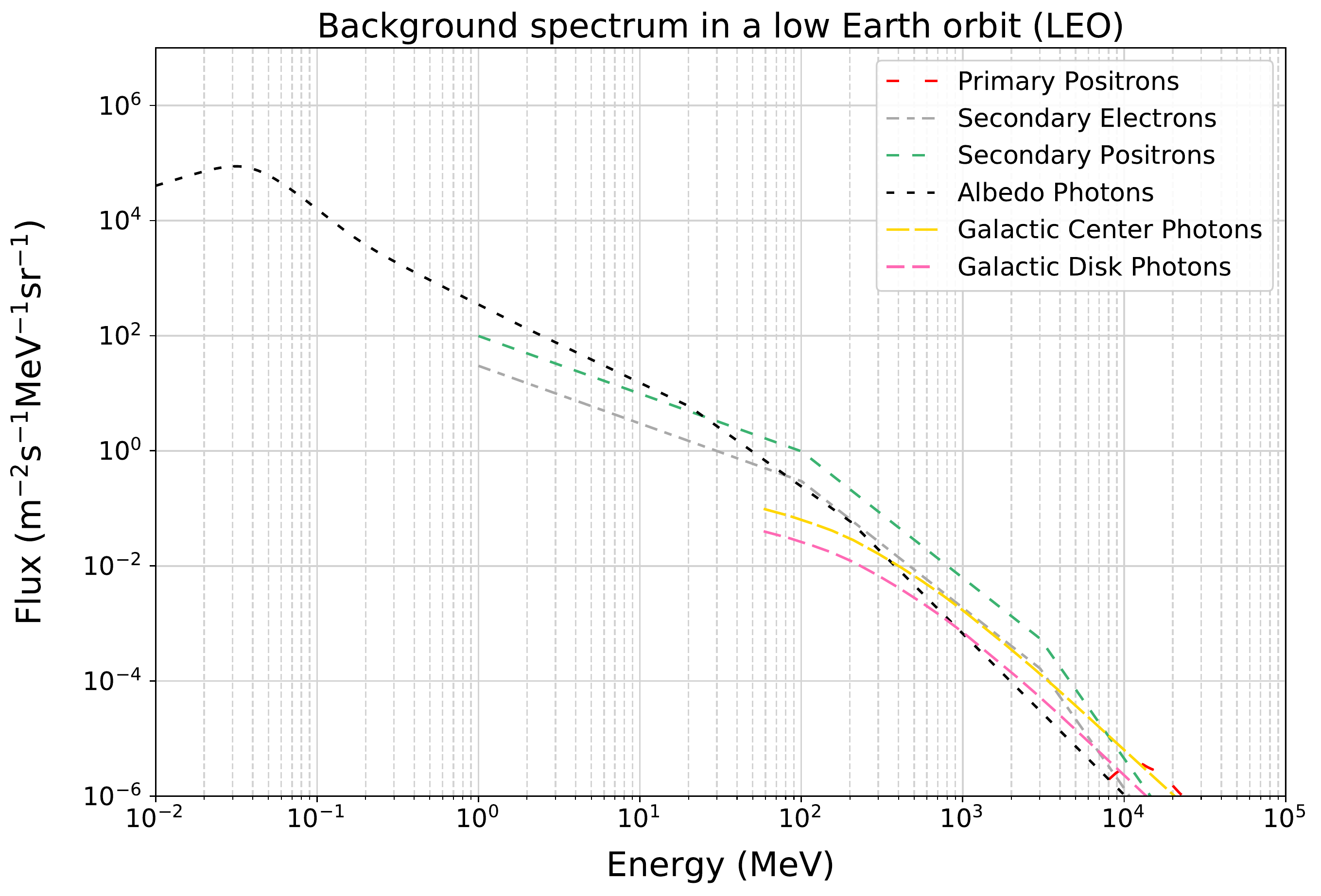}
     \caption{}
     \label{fig:spectrum_compare_part1}
  \end{subfigure}
  \begin{subfigure}[b]{0.49\textwidth}
     \centering
     \includegraphics[width=\textwidth]{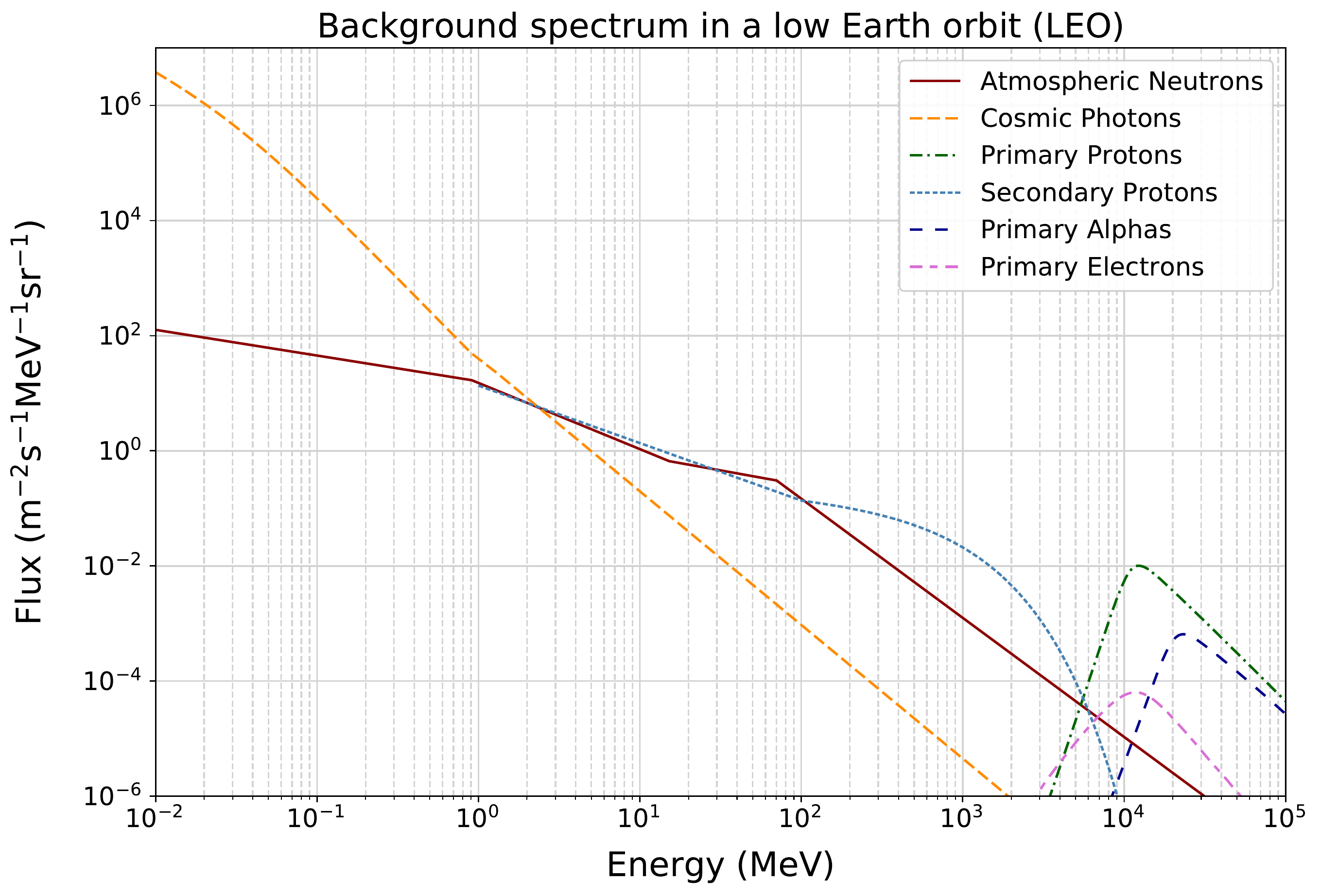}
     \caption{}
     \label{fig:spectrum_compare_part2}
  \end{subfigure}
  \begin{subfigure}[b]{\textwidth}
     \centering
     \includegraphics[width=0.50\textwidth]{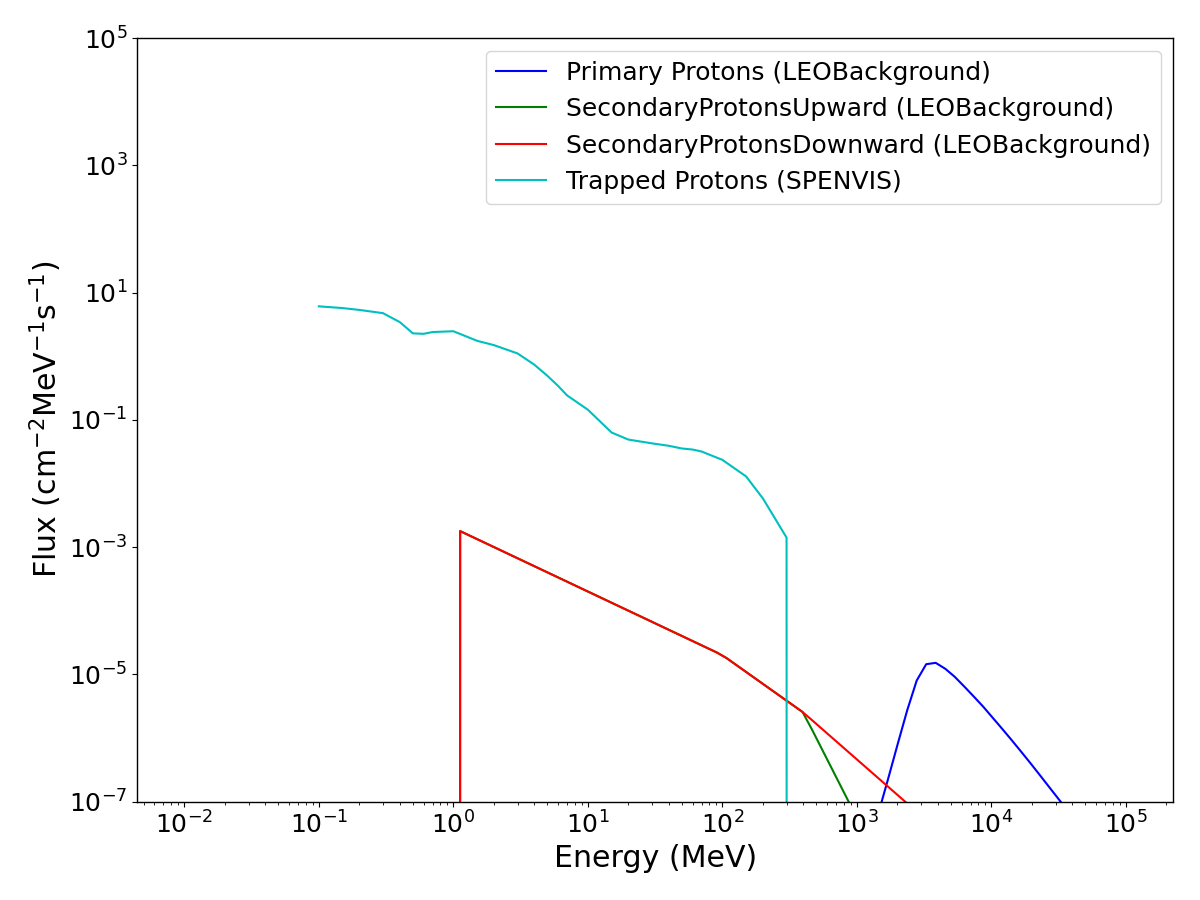}
     \caption{}
     \label{fig:proton_spectrum_compare}
  \end{subfigure}
  \hfill
  \caption{\textbf{a)} and \textbf{b)} The spectrum of various background particles \cite{leobackground_2019} expected to be seen at an altitude of \SI{383}{km} and an inclination of $42^{\circ}$ (typical POLAR-2 orbit). \textbf{c)} Comparison between the proton energy spectrum from LEOBackground \cite{leobackground_2019} and SPENVIS \cite{1989AIPC..186..483D}.  }
  \hfill
\end{figure}

As mentioned in the motivation part, it is pertinent to obtain a good estimate on radiation damage from background radiation. Therefore, the POLAR-2 simulation framework was enhanced to allow for background information to be taken from LEOBackground \cite{leobackground_2019} or SPENVIS \cite{1989AIPC..186..483D}. The former is a summary of collected data and analyses performed by various experiments; such as \textit{Fermi}-LAT \cite{LAT}, AMS-02 \cite{AMS}, etc. It provides information on neutrons, photons, protons and electrons. Figures~\ref{fig:spectrum_compare_part1} and \ref{fig:spectrum_compare_part2} show the expected flux for these particles. It is generated by setting the orbital altitude and inclination to \SI{383}{km} and $\SI{42}{^{\circ}}$ respectively. The solar modulation potential is set to \SI{650}{MV}. SPENVIS only provides information on trapped proton and electron fluxes. However, it is more comprehensive when accounting for solar activity, natural radiation belts, plasmas and gases. Protons are among the most penetrating particles (with neutrons and gammas), whereas electrons are significantly more abundant but easily stopped by the carbon fiber shielding before reaching sensitive components. Considering penetration depth, energy spectrum and deposition it is expected for protons to be the most contributing factor to deposit energy in the SiPM array and other electronics. The South Atlantic Anomaly (SAA) \cite{SAA} is a region at Earth’s surface, where the intensity of the magnetic field is particularly low. In this region solar energetic particles penetrate Earth’s atmosphere more deeply causing problems including spacecraft electronic systems. Flux of trapped particles in the SAA is at least two orders of magnitudes higher than that of primary and secondary protons (see Figures~\ref{fig:spectrum_compare_part2} and \ref{fig:proton_spectrum_compare}, where the proton spectrum is that from \cite{osti_7094195} (used by SPENVIS) and \cite{AMS} (used by LEOBackground). We therefore decide to use SPENVIS data to simulate the expected radiation \textit{dose}\footnote{Energy deposit per mass of sensitive volume.}. \newline

\begin{figure}[!h]
  \centering
  \includegraphics[width=0.7\textwidth]{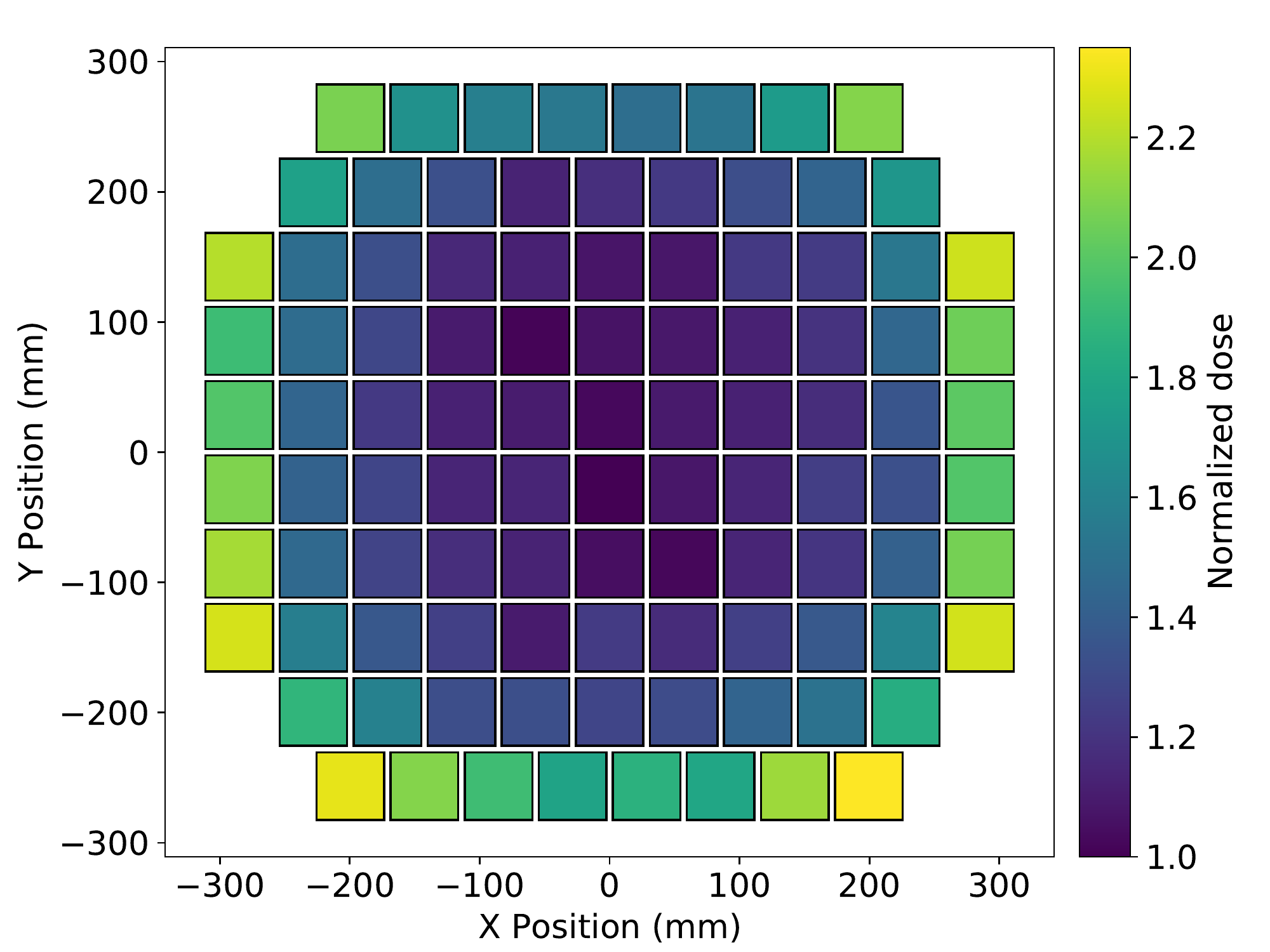}
  \caption{Normalized dose distribution per detector module in the POLAR-2 instrument (seen from the top) due to background radiation. The asymmetry is due to shielding from neighbor instruments on the payload platform and from the robotic arm adapter piece. The bottom right corner is not facing any neighbor payload, and the adapter is installed on the bottom side.}
  \label{fig:dose_density}
\end{figure}

Figure~\ref{fig:dose_density} shows, the dose (normalized with respect to the dose in the less exposed detector module) in each module. As anticipated, the dose is highest for modules closest to the edge of the instrument. For simplicity, in this paper, the 'dose' refers to the average dose expected for all POLAR-2 modules\footnote{Note that there are 100 modules, each with a 64 channel SiPM.}. In other words, the dose is defined as

\begin{equation} \label{eq_dose}
dose^{\rm POLAR-2}_{\rm SiPM} = \frac{\Sigma_{i} E^{i}_{\rm SiPM-channel}}{N_{\rm mod} \cdot N_{\rm ch} \cdot m_{\rm ch}},
\end{equation}
where $N_{mod}$, $N_{ch}$ and $m_{ch}$ refer to the number of modules (100), number of SiPM channels (64) and silicon mass of each SiPM channel. \newline

We use the AP-8 proton model \cite{osti_7094195} during a solar maximum for protons with energies ranging from $\SI{100}{keV}$ to $\SI{400}{MeV}$. Other particles (such as electrons) are excluded as they are primarily blocked by the carbon fiber shielding. The proton flux is then used to set the primary proton energy (interpreting it as an 'energy distribution'). These primaries are then randomly placed on a sphere with an isotropic distribution. Their angular distribution follows Lambert's cosine-law. To obtain the doses, \SI{35e6}{particles} were injected. Their anticipated doses are listed in the Table~\ref{tab:rad_dose} below for 4 different scenarios. The \textit{bare SiPM} reflects the scenario where the component is not shielded by neighbouring detector components and serves as the 'worst case' scenario, which is considered by us in this paper. The \textit{full instrument} reflects the full POLAR-2 instrument as shown in Figure~\ref{fig:polar2_g4}. The addition of \textit{CSS} implies that the simplified description of the China Space Station is included (top right in Figure~\ref{fig:polar2_g4}). This allows us to understand how much the shadow of the space station affects the anticipated radiation dose. As can be seen in Table \ref{tab:rad_dose}, the annual doses at three different altitudes are shown. These reflect the extremities of the altitude of the CSS (\SI{340}{km} and \SI{450}{km}) and its expected typical altitude (\SI{383}{km}). When converting the dose to "space equivalent year", the dose rate expected at \SI{383}{km} is used.

\begin{table}[ht]
\centering
\begin{tabular}{ | c | c | c | c | }
\hline
Simulation Setup & SiPM dose & SiPM dose & SiPM dose \\ 
 & per year at & per year at & per year at \\ 
 & \SI{340}{km} (Gy/yr) & \SI{383}{km} (Gy/yr) & \SI{450}{km} (Gy/yr) \\
\hline
Bare SiPM & 1.29 & 2.79 & 8.86 \\ 
\hline
Bare SiPM + CSS & 1.13 & 2.46 & 7.75 \\ 
\hline
Full instrument & $4.10 \times 10^{-2}$ & $8.24 \times 10^{-2}$ & $1.82 \times 10^{-1}$ \\ 
\hline
Full instrument + CSS & $3.89 \times 10^{-2}$ & $7.89 \times 10^{-2}$ & $1.74 \times 10^{-1}$ \\ 
\hline
\end{tabular}
\caption{Radiation doses of a SiPM for four different scenarios. The most pessimistic scenario is the 'Bare SiPM', whereas the most realistic one is the 'Full Instrument + CSS'.}
\label{tab:rad_dose}
\end{table}

\section{Experimental setup}

 All planned irradiation sessions were performed at the proton radiotherapy facility at Institute of Nuclear Physics Polish Academy of Sciences (IFJ PAN) in Krakow. The radiation campaign was supposed to reproduce the integrated dose the POLAR-2 instrument faces during its lifetime in orbit (which includes cosmic radiation when it passes through the SAA or if there are any solar events). The total dose, activation and performance of the SiPM array is of interest. For comparison, the newer series S14161-6050HS-04 (from here on referred to as S14161) from Hamamatsu was also irradiated with doses later listed in Table~\ref{tab:mppc_rad_table}.\newline 

\begin{figure}[h]
  \begin{subfigure}{0.49\textwidth}
    \centering
    \includegraphics[scale=0.17]{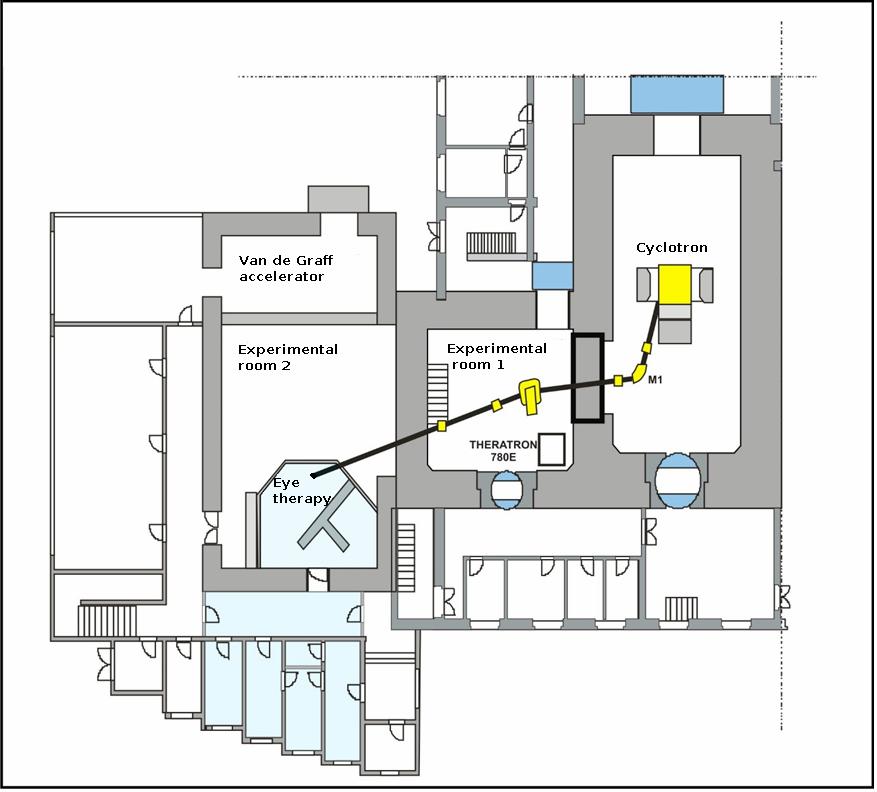}
    \caption{}
    \label{fig:IFJ_proton_facility}
    \hfill
  \end{subfigure}
  \begin{subfigure}{0.49\textwidth}
    \centering
    \includegraphics[scale=0.04]{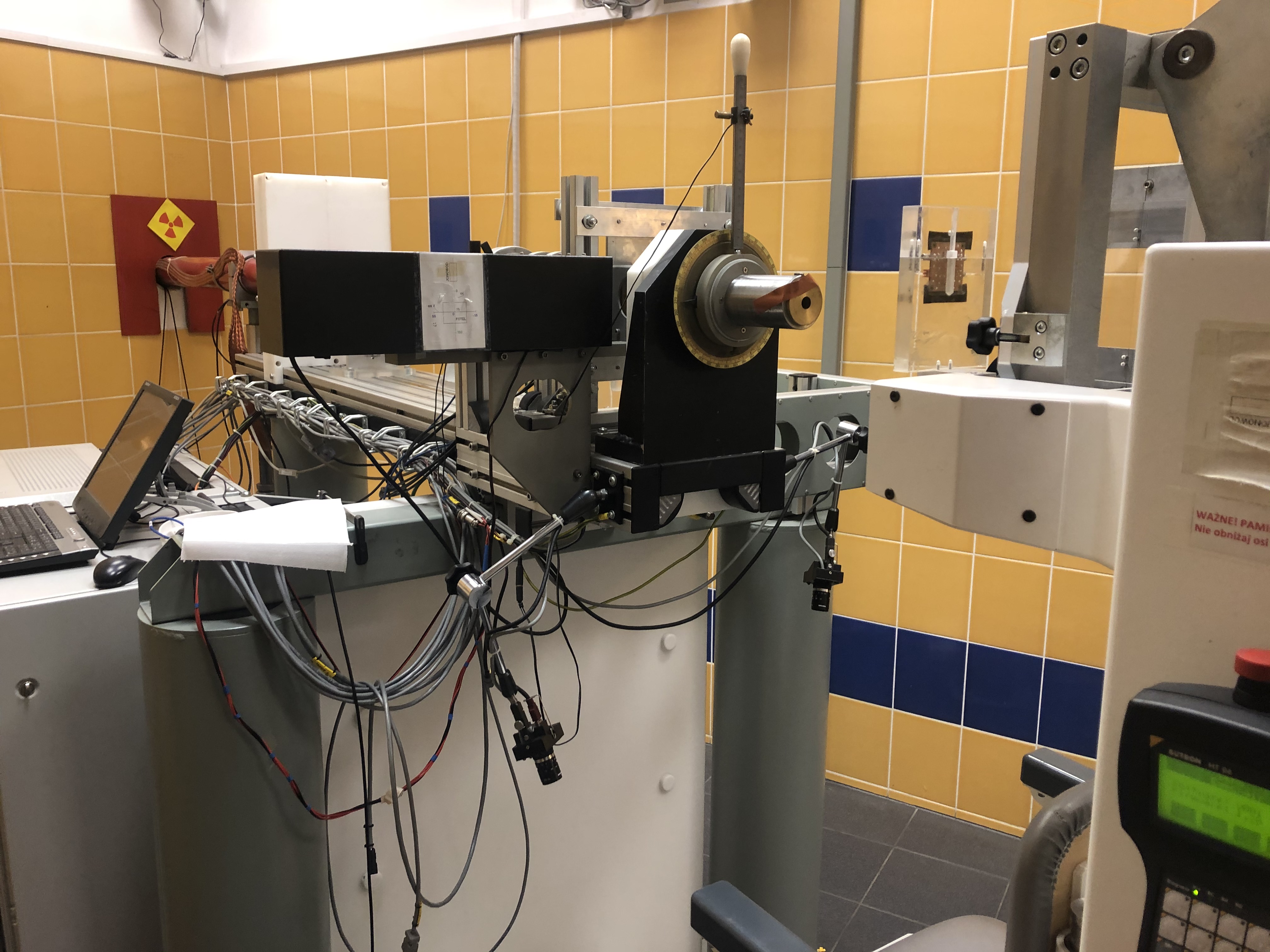}
    \caption{}
    \label{fig:IFJ_facility_photo}
    \hfill
  \end{subfigure}
  \caption{\textbf{a)} A schematic of the proton radiation therapy facility at IFJ. \textbf{b)} A photo of the exit of proton beam line.}
\end{figure}

To generate the proton beam, the particles were accelerated by the AIC-144 isochronous cyclotron facility at an energy of \SI{58}{MeV}. They were then transported from the facility to the sample by a system consisting of bending and correction magnets, measuring boxes and quadrupole doublets. It then passes through a tantalum-aluminum collimator where it scatters on a $\SI{25}{\mu m}$ tantalum proton scattering foil \cite{MICHALEC2010738}. A schematic of this facility and a photo of the exit of the beam line and setup for the samples are respectively shown in figures \ref{fig:IFJ_proton_facility} and \ref{fig:IFJ_facility_photo}. 

\begin{figure}[!h]
  \centering
  \includegraphics[width=0.7\textwidth]{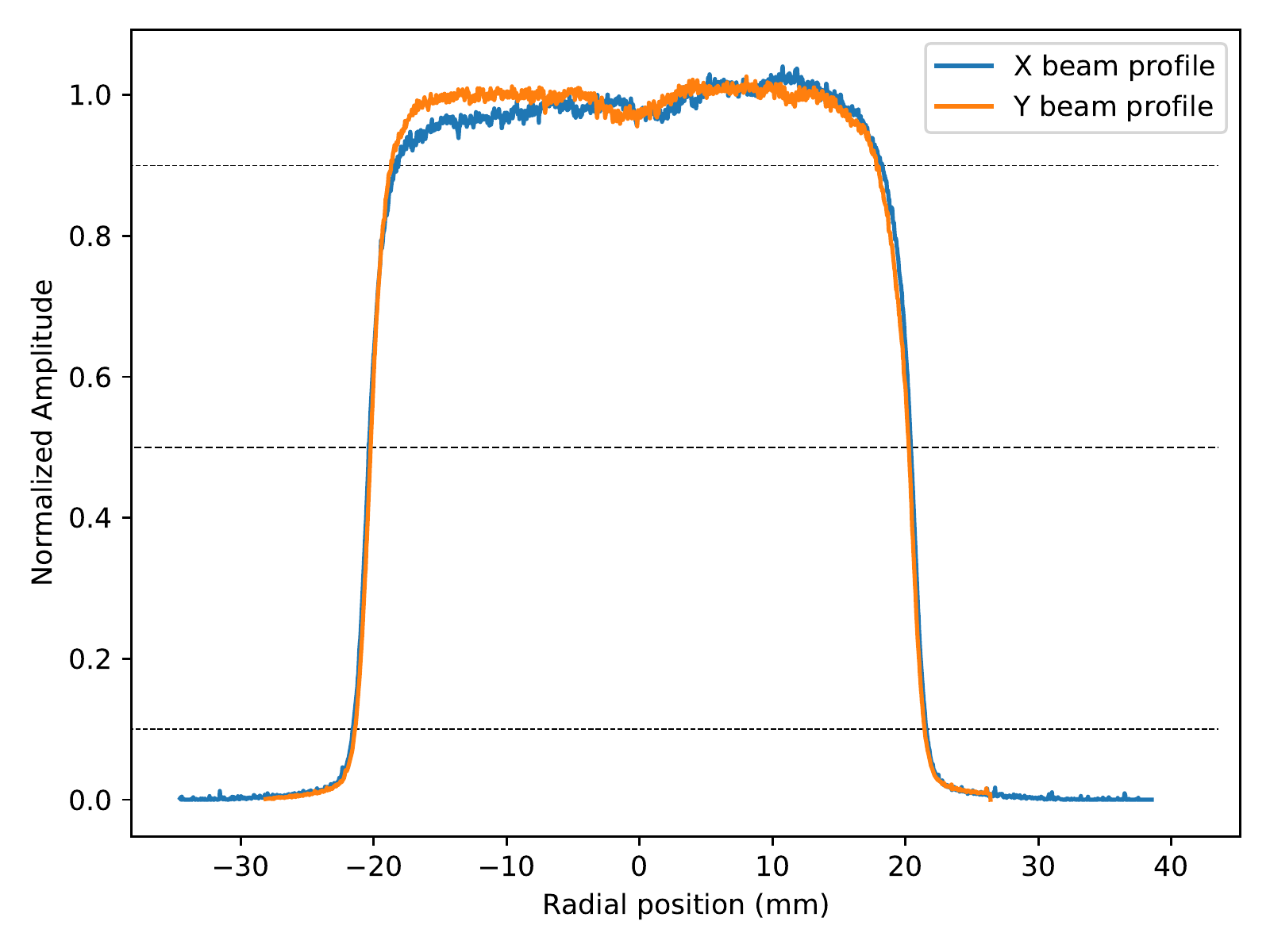}
  \caption{Scaled beam profile (to its average) of \SI{58}{MeV} protons along the x-axis and y-axis. }
  \label{fig:beam_profile}
\end{figure}

\begin{figure}[!h]
  \centering
  \includegraphics[width=0.6\textwidth]{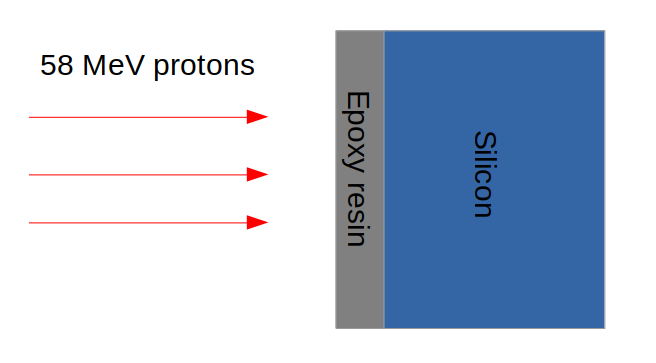}
  \caption{The model of SiPM single channel used for different dose calculations. The first layer corresponds to $\SI{100}{\mu m}$ of epoxy resin, the second layer to $\SI{450}{\mu m}$ of silicon.}
  \label{fig:single_cell}
\end{figure}

In Figure~\ref{fig:beam_profile} we show the beam profile of the \SI{58}{MeV} protons exiting the collimator (i.e. the profile which would arrive to the sample). It can be seen that the beam profile is mostly stable within the 5$\%$ level, consistent with recent work with the facility \cite{MICHALEC2010738}. Depending on the sample, a \SI{10}{mm} thick donut-shaped brass disk is placed to exclude neighbouring components from being irradiated.
To simulate the expected dose from the aforementioned beam specification, we inject \SI{1e6}{protons} perpendicular to the SiPM, as modelled in the POLAR-2 simulation framework, surface (pointing first into the resin). This yields the average dose per proton, later combined with the proton fluence to obtain the total irradiated dose. A simple cross-sectional schematic is shown in Figure~\ref{fig:single_cell}. The simulations are also performed with GEANT4. \newline

\begin{figure}[!h]
  \centering
  \includegraphics[width=0.7\textwidth]{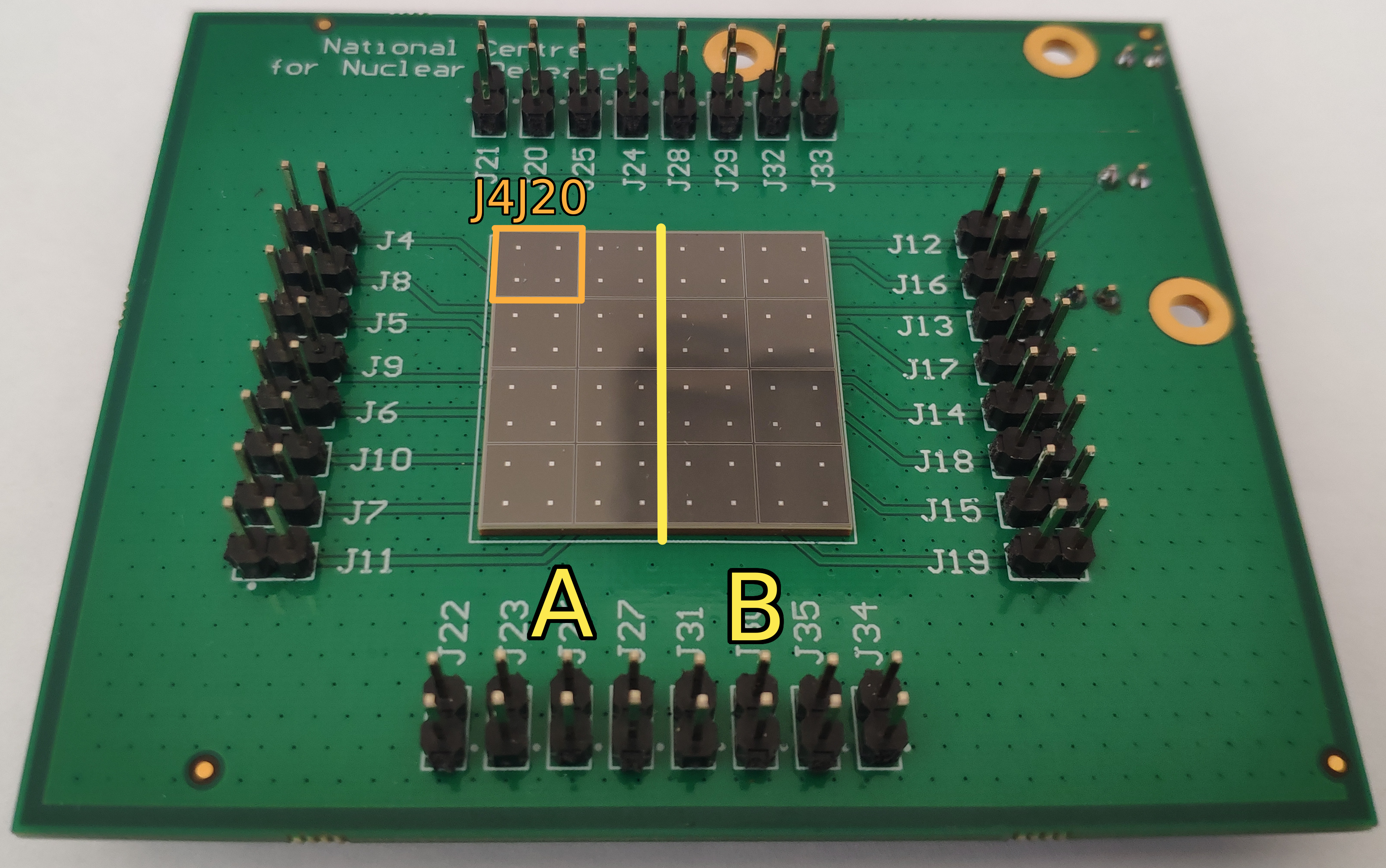}
  \caption{The Hamamatsu MPPC S13361-6075NE-04 array mounted on PCB plate. Yellow line divide the array to two sections 'A' and 'B'. Single channel J4J20 is also marked with orange line.}
  \label{fig:MPPC}
\end{figure}

A single SiPM array was divided into two sections, noted by 'A' and 'B' as shown in Figure~\ref{fig:MPPC}, to irradiate them with different fluences. Each section consists of a subarray of 2$\times$4 channels, corresponding to half of the SiPM array. The arrays are mounted on a PCB, allowing us to bias and read the output signals from a chosen single channel or whole subarray. For arrays S13361-1, S13361-2 and S14161-1, during each subarray irradiation, the second half was shielded by a $\SI{7}{mm}$ thick lead cover, fully stopping the incoming protons. The whole S14161-2 array was irradiated with the same dose, while section 'A' was biased by \SI{1}{V} of overvoltage\footnote{The excess bias voltage over the breakdown voltage, V$_{bd}$, is called overvoltage (V$_{ov}$).}. Table~\ref{tab:mppc_rad_table} lists how the samples are irradiated,their proton fluences, calculated doses and space equivalent exposure time for a 'Full Instrument + CSS' and 'Bare SiPM' scenarios. The dose in the SiPM is computed the same way as shown in Equation~\ref{eq_dose} (where $N_{modules}=1$). 

\begin{table}[ht]
\centering
\hspace*{-1.3cm}\begin{tabular}{ | c | c | c| c| c | c |}
\hline
MPPC & Fluence & Total Dose & NIEL Dose & \multicolumn{2}{|c|}{Equivalent Years in Space for POLAR-2}\\ 
Type &($\frac{protons}{cm^2}$) & (Gy) & (mGy) & 'Full Instr. + CSS' & 'Bare SiPM' \\ 
\hline
S13361-1 sec. A  & $2.00 \times 10^8$ & 0.267 & 0.174 & 3.38  & $9.57 \times 10^{-2}$ \\ 
\hline
S13361-1 sec. B  & $6.10 \times 10^8$ & 0.815 & 0.532 & 10.3 & $2.92 \times 10^{-1}$ \\
\hline
S13361-2 sec. A  & $1.64 \times 10^9$ & 2.19 & 1.43 & 27.8  & $7.85 \times 10^{-1}$\\ 
\hline
S13361-2 sec. B  & $3.73 \times 10^9$ & 4.96 & 3.24 & 62.9 & 1.78 \\
\hline
S14161-1 sec. A & $1.90 \times 10^8$ & 0.254 & 0.166 & 3.22 & $9.10 \times 10^{-2}$ \\
\hline
S14161-1 sec. B & $1.73 \times 10^9$ & 2.31 & 1.51 & 29.3 & $8.28 \times 10^{-1}$ \\
\hline
S14161-2 sec. A and B  & $1.73 \times 10^9$ & 2.31 & 1.51 & 29.3 & $8.28 \times 10^{-1}$ \\
\hline

\end{tabular}
\caption{An overview of SiPM elements irradiated, their corresponding dose and equivalent time in space for a 'Full Instrument + CSS' and 'Bare SiPM' scenarios (as described in section \ref{sec:leo_bkg}).}
\label{tab:mppc_rad_table}
\end{table}

\section{Experimental results}

In order to fully characterise the behavior of SiPMs to radiation damage, the current-voltage characteristics and dark count spectra have been analysed before and after its exposure to the proton beam. Their activation products have also been studied to understand any long-term effects. \newline

\subsection{Current-voltage characteristics}

All current-voltage (I-V) characteristics were measured before and after each successive irradiation session using the Keysight 2900B Source Measurement Unit. The voltage scale was chosen over the range of 40-\SI{60}{V} for the S13361 series and 30-\SI{50}{V} for the S14161 series in order to cover a wide range of V$_{ov}$. The current readout was limited to $\SI{100}{\mu A}$ before irradiation and increased to $\SI{200}{\mu A}$ for cases after irradiation. Based on the Keysight 2900B specification and chosen current range the current resolution was limited to about \SI{60}{nA}. \newline

I-V curve measurements were taken \SI{2}{h}, \SI{26}{h}, \SI{8}{days} and \SI{2}{months} after irradiation to study potential annealing effects. The only exception to this is, SiPM S13361-2, where I-V characteristics were not measured \SI{2}{h} after irradiation. It was first taken to a HPGe detector to measure its activation (see later section \ref{sec:activation}).
The former two I-V curves were taken at the IFJ experimental hall where the room temperature is 27$\pm\SI{0.8}{^{\circ}C}$. After completion of the radiation campaign, the SiPMs were placed inside a climate chamber where the temperature is set to $\SI{25}{^{\circ}C}$. The $\SI{2}{^{\circ}C}$ temperature difference corresponds to a bias voltage (or overvoltage) change of about \SI{0.11}{V} for the S13361 series and \SI{0.07}{V} for the S14161 series (based on Hamamatsu specification: \SI{54}{\frac{mV}{^{\circ}C}} and \SI{34}{\frac{mV}{^{\circ}C}} respectively). The breakdown voltage describes the minimum bias voltage required (in reverse direction) which leads to self-sustaining avalanche multiplication. In other words, it is the minimum bias for which the pulses stay visible at the SiPM output. Our studies show, that the standard deviation of $V_{bd}$ between different channels for a single array is in the range of \SI{0.12}{V}. As the change in bias voltage is within 1$\sigma$, we conclude that the temperature difference is small enough for a direct comparison of the I-V characteristics. \newline

\begin{figure}[!h]
  \centering
  \begin{subfigure}[b]{0.4\textwidth}
    \centering
    \includegraphics[width=\textwidth]{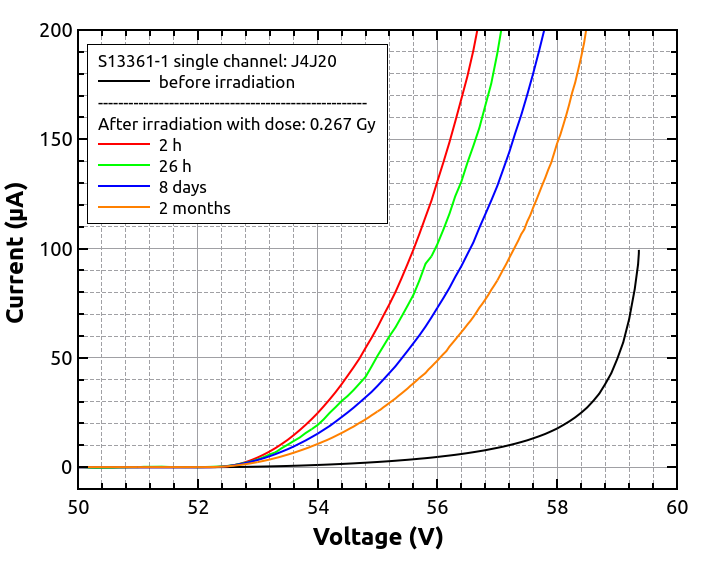}
    \caption{}
    \label{fig:S13_IV_0.4Gy}
  \end{subfigure}
  \begin{subfigure}[b]{0.4\textwidth}
    \centering
    \includegraphics[width=\textwidth]{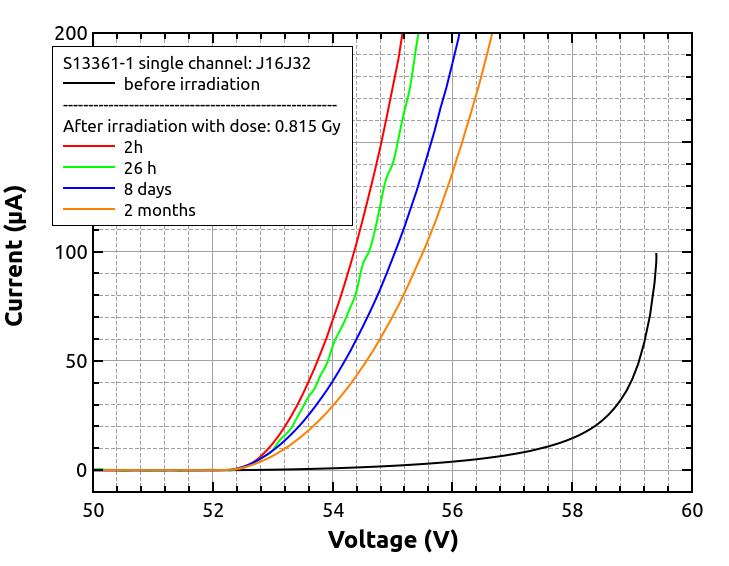}
    \caption{}
    \label{fig:S13_IV_1.2Gy}
  \end{subfigure}
    \begin{subfigure}[b]{0.39\textwidth}
    \centering
    \includegraphics[width=\textwidth]{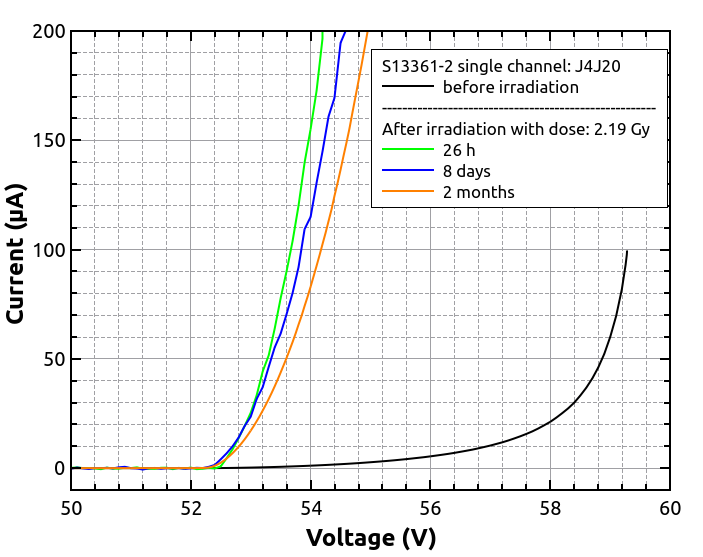}
    \caption{}
    \label{fig:S13_IV_3.5Gy}
  \end{subfigure}
  \begin{subfigure}[b]{0.39\textwidth}
    \centering
    \includegraphics[width=\textwidth]{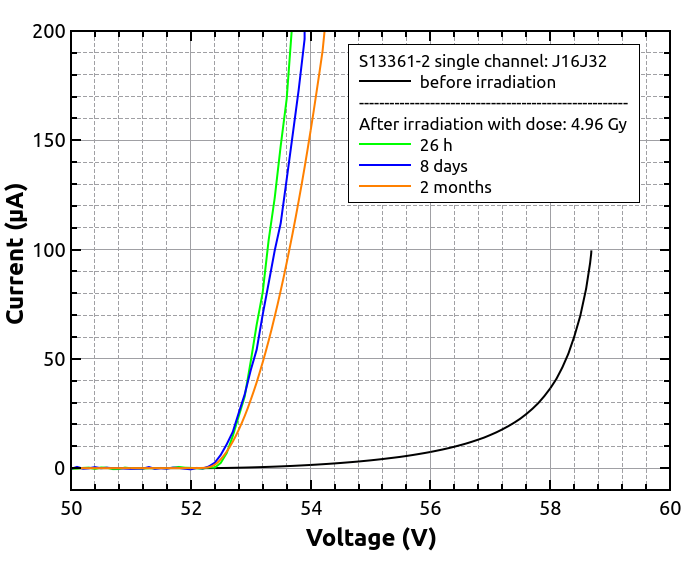}
    \caption{}
    \label{fig:S13_IV_7.0Gy}
  \end{subfigure}
  \caption{Current-voltage S13361 single channel characteristics before and after its irradiation for a dose of \textbf{a)} \SI{0.267}{Gy}, \textbf{b)} \SI{0.815}{Gy}, \textbf{c)} \SI{2.19}{Gy} and \textbf{d)} \SI{4.96}{Gy}.}
  \hfill
  \label{fig:S13361_IV}
\end{figure}

\begin{figure}[!h]
  \centering
  \begin{subfigure}[b]{0.4\textwidth}
    \centering
    \includegraphics[width=\textwidth]{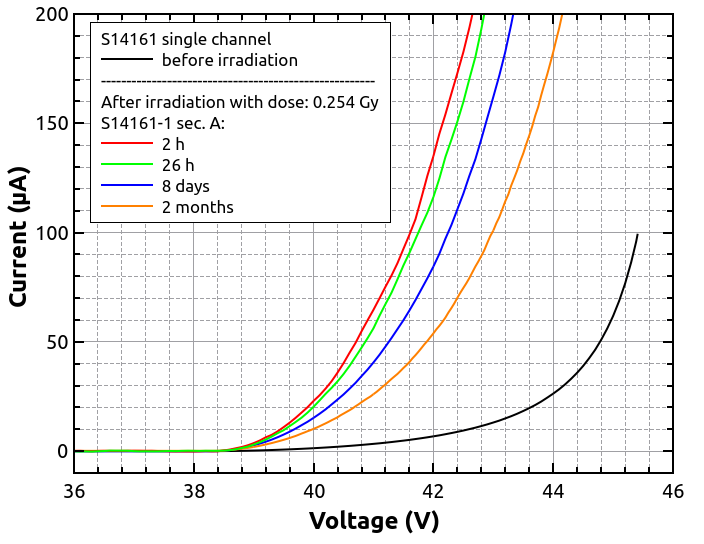}
    \caption{}
    \label{fig:S14_IV_0.4Gy}
  \end{subfigure}
  \begin{subfigure}[b]{0.4\textwidth}
    \centering
    \includegraphics[width=\textwidth]{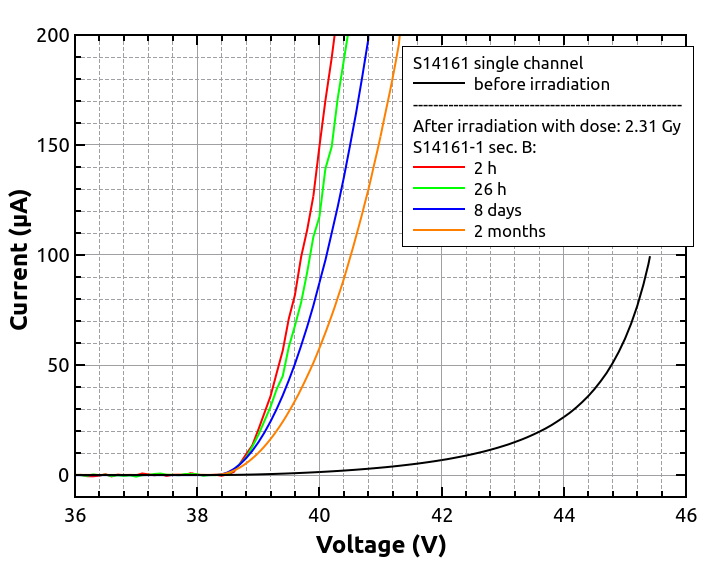}
    \caption{}
    \label{fig:S14_IV_3.5Gy}
  \end{subfigure}
  \caption{Current-voltage MPPC S14161 single channel characteristics before and after its irradiation for a dose of \textbf{a)} \SI{0.254}{Gy} and \textbf{b)} \SI{2.31}{Gy}.}
  \hfill
  \label{fig:S14161_IV}
\end{figure}

Figure~\ref{fig:S13361_IV} shows the obtained I-V characteristics for the S13361 SiPM series for different doses and time before and after irradiation. The I-V characteristics for the S14161 MPPC series are presented in Figure~\ref{fig:S14161_IV}. It is evident, when exposed to the same dose, that the current rate increase is higher for S14161 than for the S13361 series. This already implies that the S13361 type is more beneficial for POLAR-2 purposes. For both SiPM types we also observed self-annealing processes, manifesting itself through a decreasing current with respect to time after its irradiation. Further details on the annealing process of SiPM arrays will be discussed in another paper. 
\newline

\begin{figure}[!h]
  \centering
  \begin{subfigure}[b]{0.4\textwidth}
    \centering
    \includegraphics[width=\textwidth]{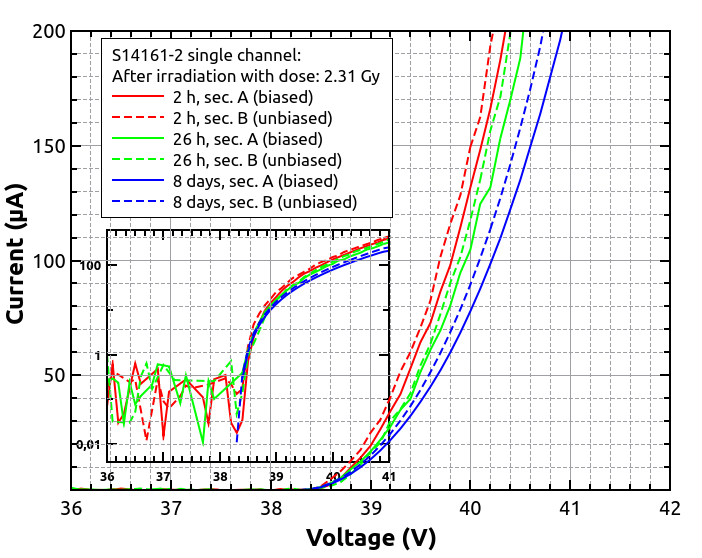}
    \caption{}
    \label{fig:S14_IV_3.5Gy_biased}
  \end{subfigure}
  \begin{subfigure}[b]{0.4\textwidth}
    \centering
    \includegraphics[width=\textwidth]{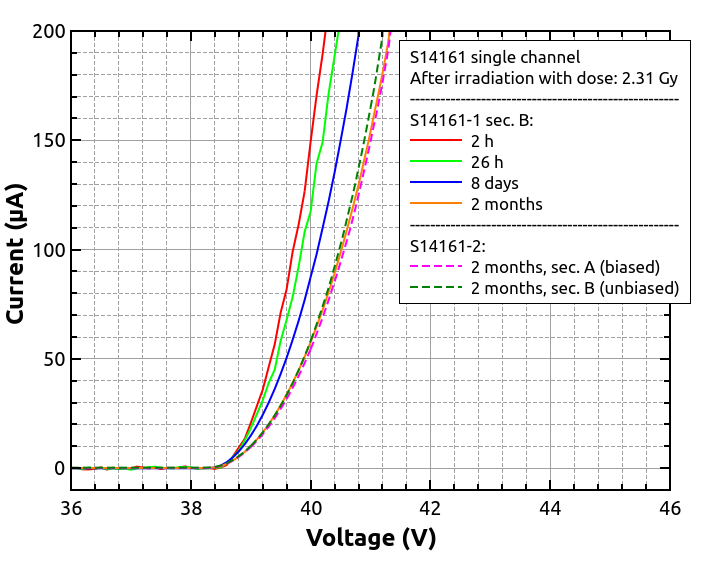}
    \caption{}
    \label{fig:S14_IV_3.5Gy_biased2}
  \end{subfigure}
  \caption{Current-voltage single channel characteristics for biased (channel: J4J20) and unbiased (channel: J16J32) S14161 subarray irradiated with dose \SI{2.31}{Gy} for different times after irradiation. The inset in figure \textbf{a)} shows the breakdown voltage region in y-log scale. Figure \textbf{b)} compares also two S14161 arrays (two single channels unbiased, one biased) two months after irradiation.}
  \hfill
  \label{fig:S14161-2_biased}
\end{figure}

To discern contributions of the bias voltage on the radiation damage (through a change in the I-V characteristics or dark count rates), the S14161-2 array was irradiated with a single dose of \SI{2.31}{Gy}, where only subarray 'A' was biased with an overvoltage of  \SI{1}{V} (while subarray 'B' remained unbiased). Figure~\ref{fig:S14161-2_biased} presents the obtained I-Vs for different times after irradiation. The inset in Figure~\ref{fig:S14_IV_3.5Gy_biased} also shows the breakdown voltage region in logarithmic scale. Taking into account differences between the breakdown voltages for single channels, we did not observe any significant bias-dependent features. Furthermore, applying a bias voltage did not affect the self-annealing process at room temperature. \newline     

\newpage

\subsection{Dark Counts analysis}

Changes in the I-V characteristics registered after proton irradiation are also reflected in the number of dark counts (DC) and the spectral shape of the output signal. To determine these parameters we adopted a method based on raw pulses analysis, originally presented by \cite{Piemonte_2012} and later applied by \cite{Otte_2017}. The chosen procedure eliminates the problem of high dead time for analogue and even digital data acquisition systems (like digitizers), where a few million pulses per second are expected. In our case, the direct signal from a single SiPM channel was additionally amplified by a transimpedance amplifier and then sent to a digital oscilloscope of type Keysight DSOS104-A (which has a sampling rate of 1 GSamples/s and a bandwidth of \SI{500}{MHz}), where \SI{10}{ms} waveforms with many single raw pulses were saved on a hard drive. In this case, each waveform was triggered based on a 0.5 p.e. threshold. Five waveforms were saved for each overvoltage and temperature, giving us an estimation of statistical uncertainties for a chosen \SI{10}{ms} period time. \newline

For the clarity, we listed below the main steps made during the off-line analysis:
\begin{itemize}
\item The Savitzky-Golay filter \cite{Sav-Gol} was applied twice to each registered waveform. This filter allows to extract p.e. peaks in a DC spectrum for lower temperature ranges. Its implementation has proven to be especially useful for the S14161, which is characterized by a smaller amplitude of 1 p.e. and a longer tail; 
\item A copy of \SI{10}{ms} waveform was created, delayed and subtracted from the original one;
\item The peaks above certain thresholds (\SI{12.5}{mV} for S13361, \SI{1.6}{mV} for S14161) were taken into account; 
\item The amplitude of each peak and the time difference between two consecutive pulses were determined. 
\end{itemize}

The python tool \textit{find$\_$peak} from the \textit{scipy~1.3.3-3} library \cite{scipy} was used for the last two items. The advantage of this procedure is its speed and pile-up sensitivity, which can be controlled by \textit{find$\_$peak} input parameters. \newline

\begin{figure}[!h]
  \centering
  \begin{subfigure}[b]{0.6\textwidth}
    \centering
    \includegraphics[width=\textwidth]{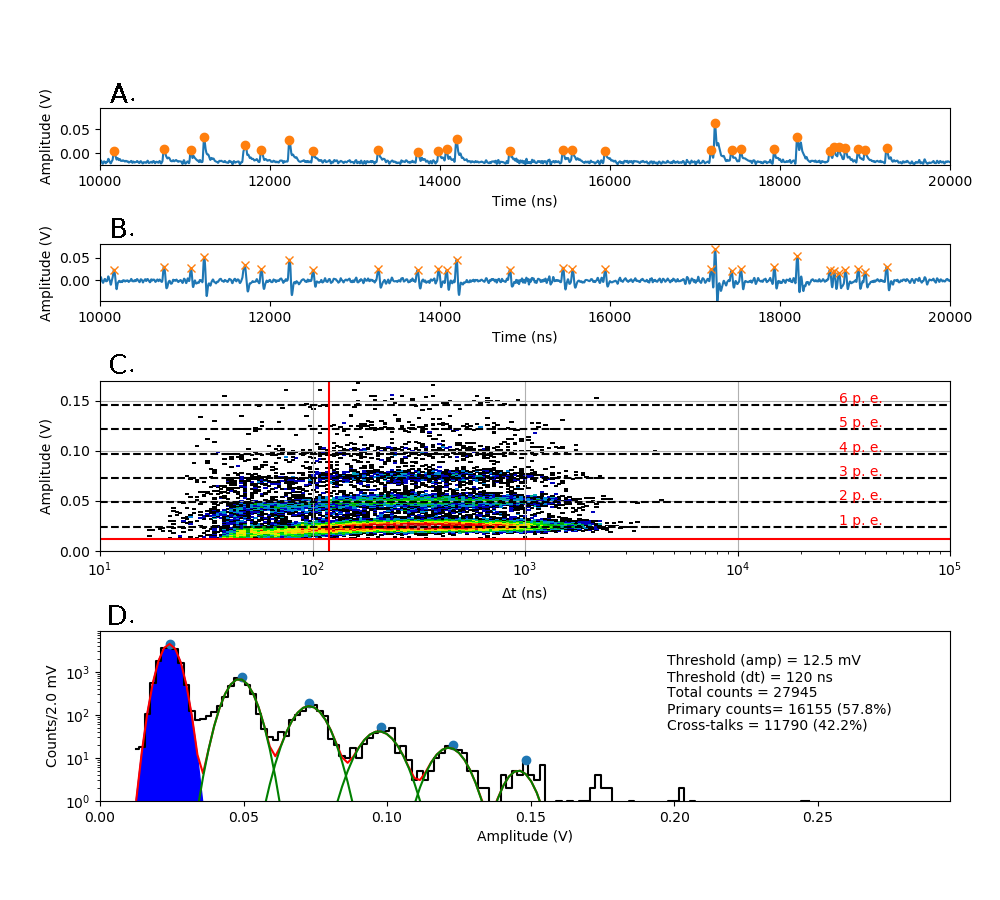}
    \caption{}
    \label{fig:S13_dcr_example}
  \end{subfigure}
  \begin{subfigure}[b]{0.6\textwidth}
    \centering
    \includegraphics[width=\textwidth]{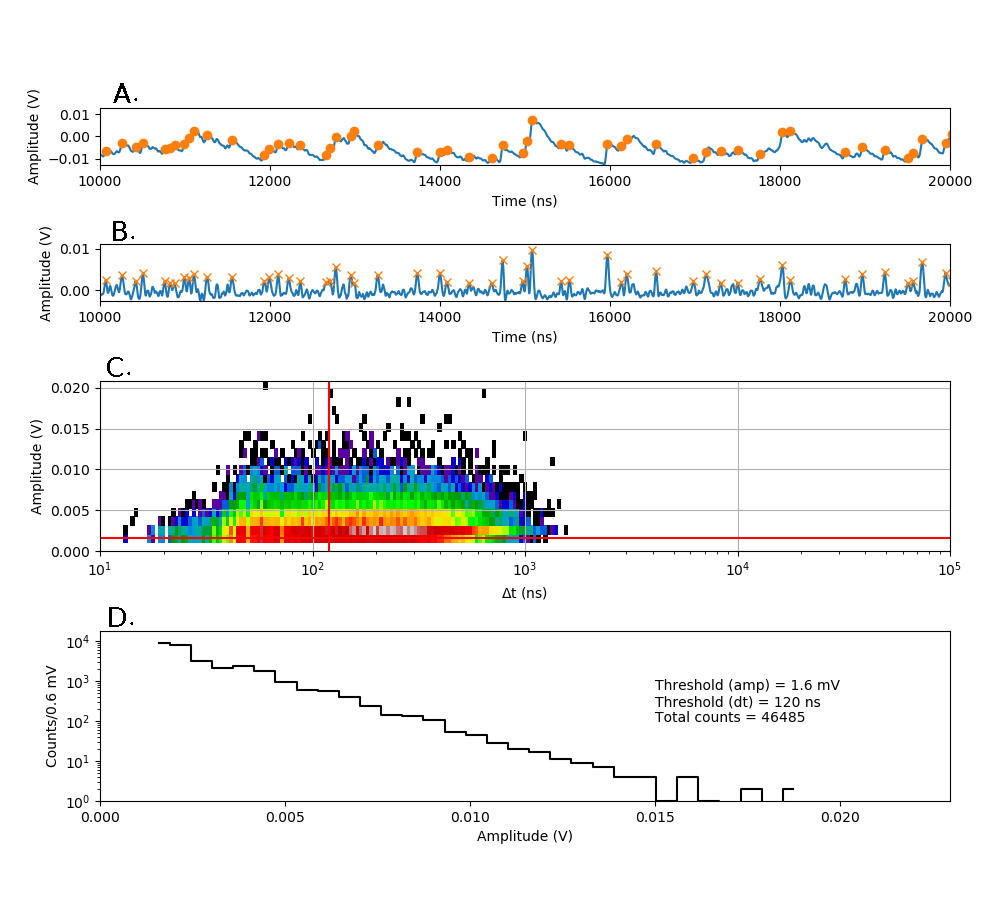}
    \caption{}
    \label{fig:S14_dcr_example}
  \end{subfigure}
  \caption{An example of a single channel waveform analysis of a) S13361 b) S14161, measured at temperature $\SI{25}{^{\circ}C}$ before proton irradiation. A - a part of \SI{10}{ms} waveform, B - transformed waveform with peak identification, C - 3D distribution of dark pulse amplitudes as a function of time difference between two subsequent events, where the Z-axis is represented by a colour scale. D - projection of the dark counts with amplitude and time cut conditions. Blue color describes 1 p.e. position and primary counts (not visible in the S14161 case).}
  \hfill
  \label{fig:S13_S14_DCR_examle}
\end{figure}

An example of the aforementioned analysis steps for a non-irradiated S13361 single channel is presented in Figure~\ref{fig:S13_dcr_example}. Measurements were performed at $\SI{25}{^{\circ}C}$ at a voltage of \SI{55.5}{V} (V$_{ov}$=\SI{3.5}{V}). The first two upper plots show a fraction of the \SI{10}{ms} waveform with many p.e. peaks and the result of the delayed waveform subtraction. Orange dots and crosses show identified peaks. The pile-up cases are also visible. The third subplot (scatter plot) presents the multiplicity of amplitudes in the unit of volts (or number of p.e.) on the y-axis and the time difference between two subsequent events on the x-axis. Finally, the bottom subplot shows the resulting DC spectrum when the amplitude and time thresholds have been applied. The structure of p.e. peaks is clearly visible in this case. A fitted multi-Gaussian function is also shown (red line), where the parametrisation was chosen to keep constant distance (gain) between the 'n'$^{th}$ and 'n-1'$^{th}$ peak. Based on the discussion presented in \cite{Piemonte_2012}, only the cases from the first p.e. peak (blue area) are classified as a primary counts. Similar data for the S14161 are presented in Figure~\ref{fig:S14_dcr_example}. Contrary to S13361, the p.e. peaks are not visible in the DC spectrum at $\SI{25}{^{\circ}C}$. \newline

\newpage

\subsubsection{Dark Counts before irradiation} \label{sec:dc_before_irradiation}

\begin{figure}[!h]
  \centering
  \begin{subfigure}[b]{0.45\textwidth}
    \centering
    \includegraphics[width=\textwidth]{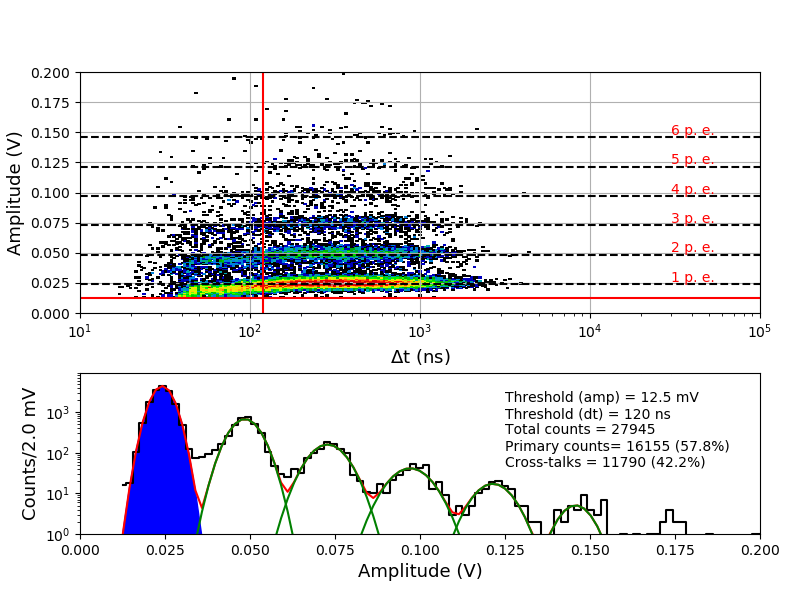}
    \caption{}
    \label{fig:S13_before_25}
  \end{subfigure}
  \begin{subfigure}[b]{0.45\textwidth}
    \centering
    \includegraphics[width=\textwidth]{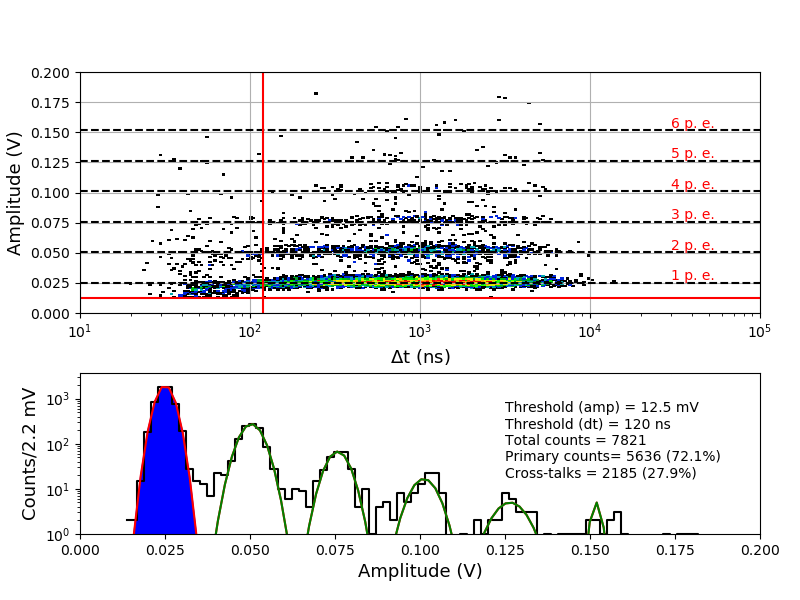}
    \caption{}
    \label{fig:S13_before_10}
  \end{subfigure}
    \begin{subfigure}[b]{0.45\textwidth}
    \centering
    \includegraphics[width=\textwidth]{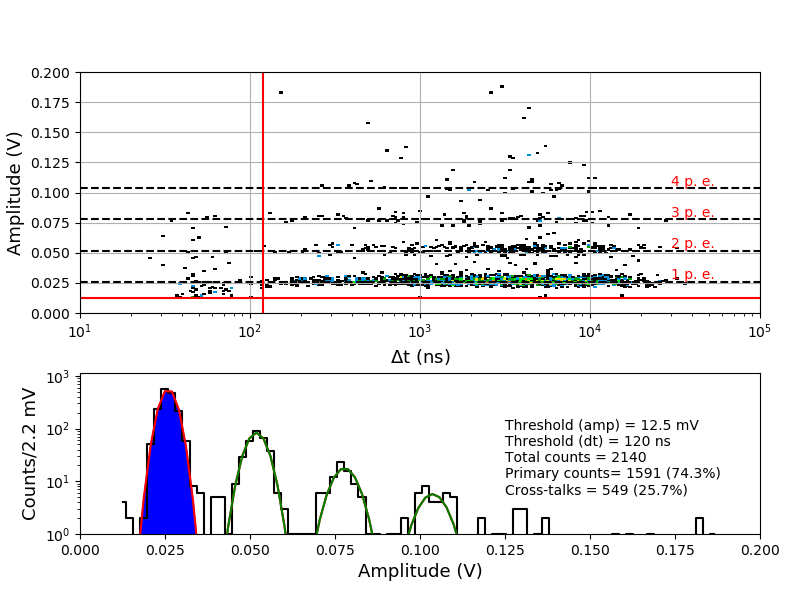}
    \caption{}
    \label{fig:S13_before_m5}
  \end{subfigure}
  \begin{subfigure}[b]{0.45\textwidth}
    \centering
    \includegraphics[width=\textwidth]{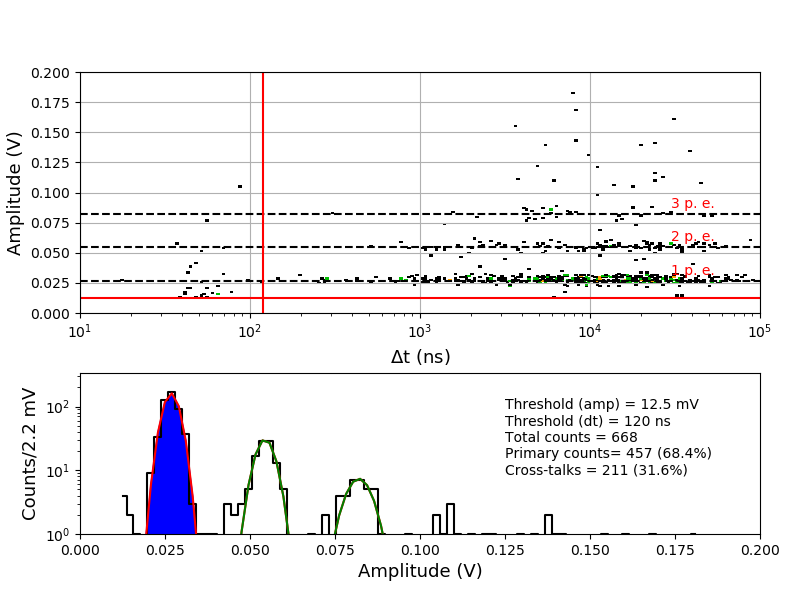}
    \caption{}
    \label{fig:S13_before_m20}
  \end{subfigure}
  \caption{S13361 single channel DC spectra measured before proton irradiation for various temperatures and the same V$_{ov}$=\SI{3.5}{V}: a) 25$^\circ$C, b) 10$^\circ$C, c) -5$^\circ$C, d) -20$^\circ$C. The Z-axis is represented here by a colour scale.}
  \hfill
  \label{fig:S13_temp_before}
\end{figure}

\begin{figure}[!hb]
  \centering
  \begin{subfigure}[b]{0.45\textwidth}
    \centering
    \includegraphics[width=\textwidth]{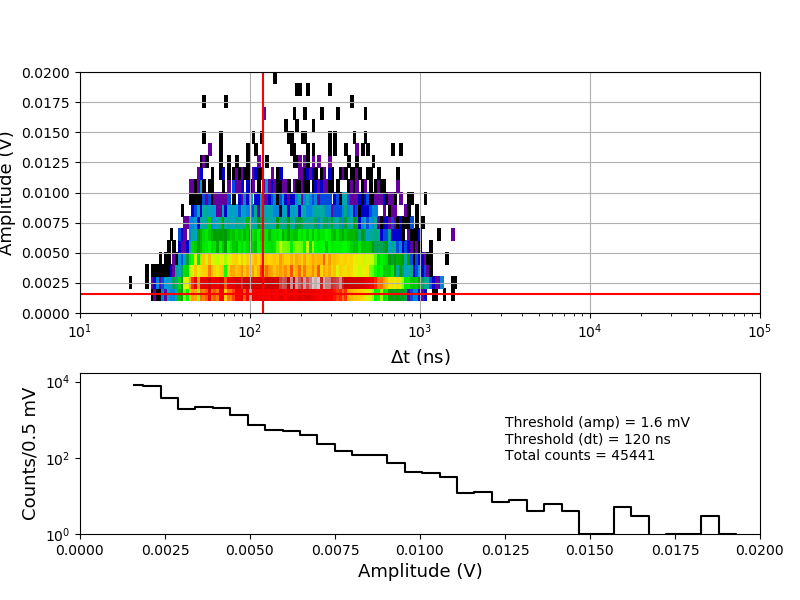}
    \caption{}
    \label{fig:S14_before_25}
  \end{subfigure}
  \begin{subfigure}[b]{0.45\textwidth}
    \centering
    \includegraphics[width=\textwidth]{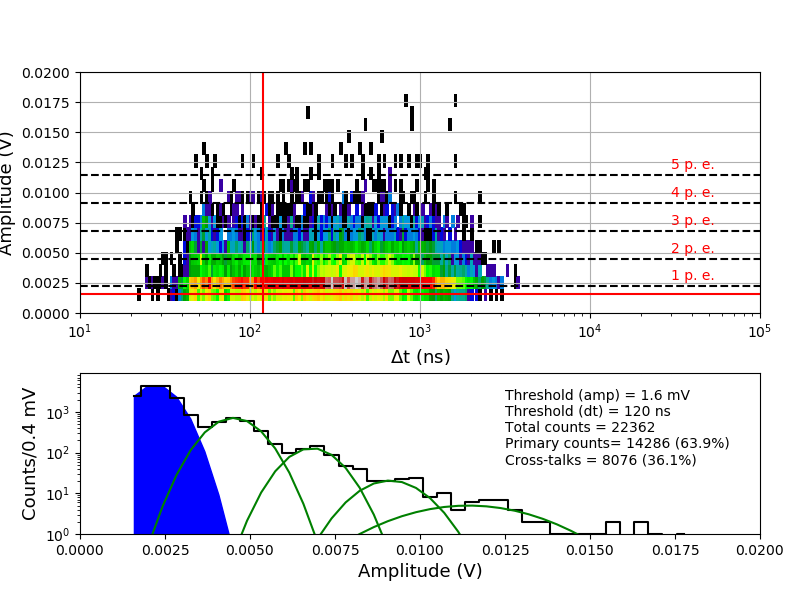}
    \caption{}
    \label{fig:S14_before_10}
  \end{subfigure}
    \begin{subfigure}[b]{0.45\textwidth}
    \centering
    \includegraphics[width=\textwidth]{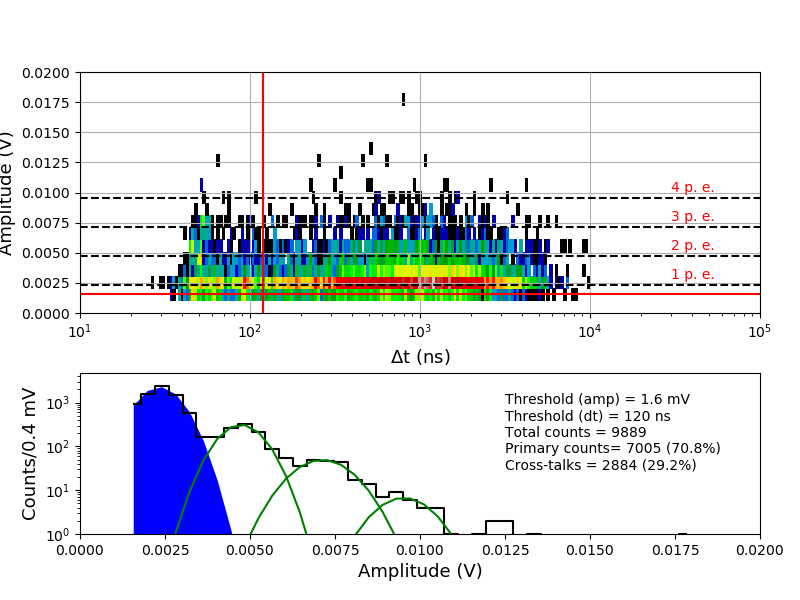}
    \caption{}
    \label{fig:S14_before_m5}
  \end{subfigure}
  \begin{subfigure}[b]{0.45\textwidth}
    \centering
    \includegraphics[width=\textwidth]{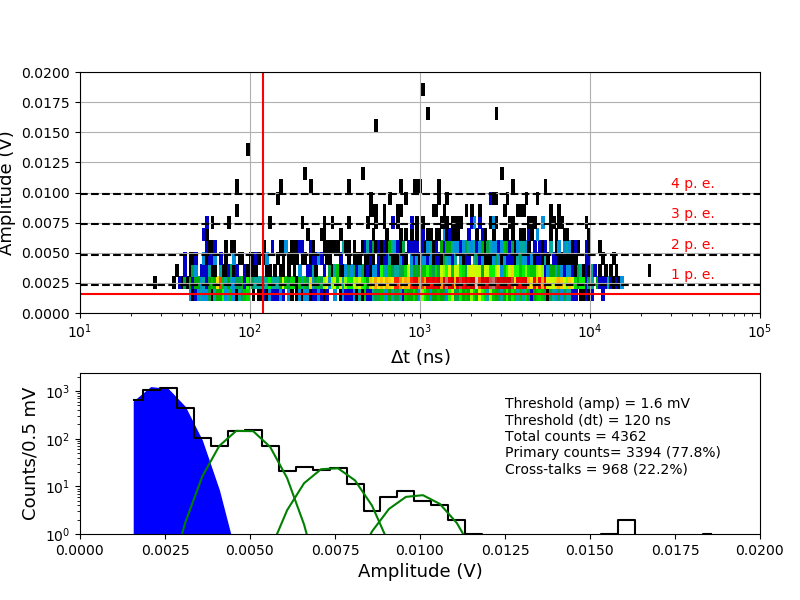}
    \caption{}
    \label{fig:S14_before_m20}
  \end{subfigure}
  \caption{S14161 single channel DC spectra measured before proton irradiation for various temperatures and the same V$_{ov}$=\SI{3.5}{V}: a) $\SI{25}{^{\circ}C}$, b) $\SI{10}{^{\circ}C}$, c) $\SI{-5}{^{\circ}C}$, d) $\SI{-20}{^{\circ}C}$. The Z-axis is represented here by a colour scale.}
  \hfill
  \label{fig:S14_temp_before}
\end{figure}

Data taken before SiPM irradiation for both series are presented below. All \SI{10}{ms} waveforms were collected in controlled temperature conditions inside a climate chamber at temperatures: $\SI{25}{^{\circ}C}$, $\SI{10}{^{\circ}C}$, $\SI{-5}{^{\circ}C}$ and $\SI{-20}{^{\circ}C}$.
Figure~\ref{fig:S13_temp_before} shows the typical DC spectra collected for S13361 with V$_{ov}$=\SI{3.5}{V}. For comparison, Figure~\ref{fig:S14_temp_before} shows the case of S14161 with the same overvoltage and temperature range. The number of total counts, primary counts and cross-talks per \SI{10}{ms} interval are also presented. As was mentioned before, the primary counts (blue area) were filtered using the same amplitude and time thresholds (about 0.5 p.e  and \SI{120}{ns} respectively). For the cross-talk it was assumed that they correspond to the total number of counts minus primary counts. No after-pulses could be distinguished, suggesting that its contributions is negligible from a statistical point of view (in this work). \newline

\begin{figure}[!t]
  \centering
  \begin{subfigure}[b]{0.45\textwidth}
    \centering
    \includegraphics[width=\textwidth]{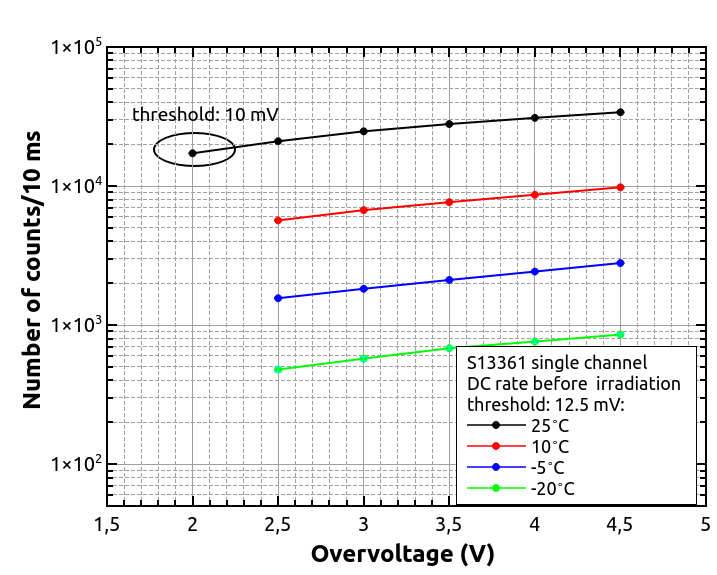}
    \caption{}
    \label{fig:S13_before_temp}
  \end{subfigure}
  \begin{subfigure}[b]{0.45\textwidth}
    \centering
    \includegraphics[width=\textwidth]{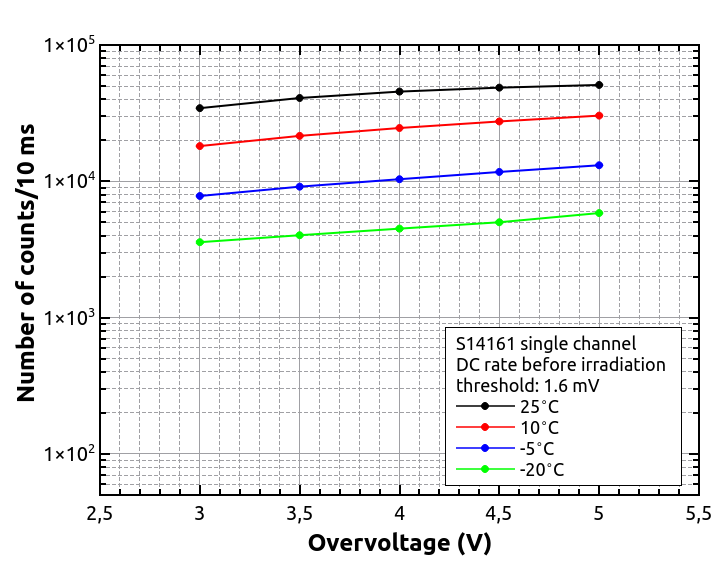}
    \caption{}
    \label{fig:S14_before_temp}
  \end{subfigure}
  \caption{The number of dark counts measured for a single channel of a) S13361 and b) S14161 array for chosen overvoltage ranges at different temperatures.}
  \hfill
  \label{fig:S13_S14_DCR_before}
\end{figure}

Summarized data for both series and chosen overvoltage ranges before proton irradiation are presented in Figure~\ref{fig:S13_S14_DCR_before}. To stress the differences between these series, the same y-axis scale was kept.
The figure shows that the DC rate is about 60$\%$ higher at $\SI{25}{^\circ C}$ for S14161. This difference increases by a factor of 4.9 for $\SI{-20}{^\circ C}$. The separation between p.e. peaks is also better for the S13361 for all temperature ranges. The extraction of single p.e. peak for S14161 at $\SI{25}{^\circ C}$ is practically impossible. \newline

\newpage
\subsubsection{Dark Counts after irradiation}

\begin{figure}[!h]
  \centering
  \begin{subfigure}[b]{0.45\textwidth}
    \centering
    \includegraphics[width=\textwidth]{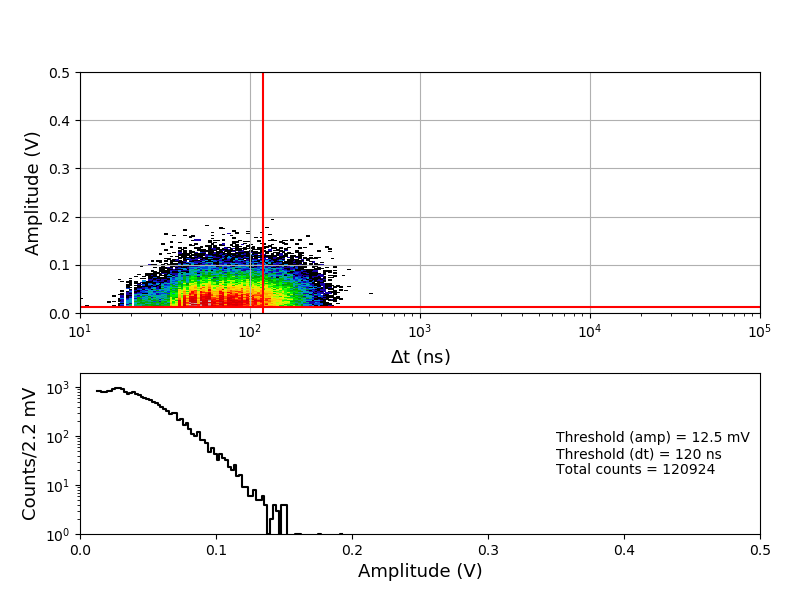}
    \caption{}
    \label{fig:S13_after_25}
  \end{subfigure}
  \begin{subfigure}[b]{0.45\textwidth}
    \centering
    \includegraphics[width=\textwidth]{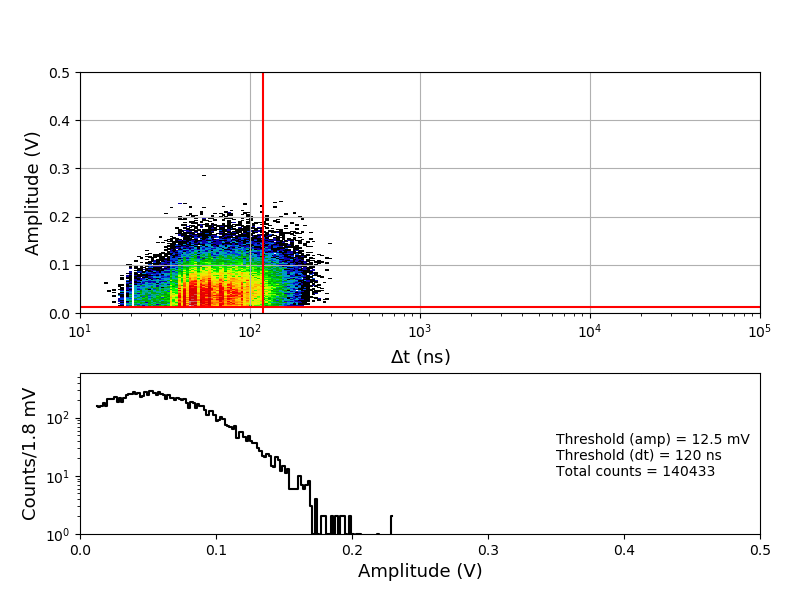}
    \caption{}
    \label{fig:S13_after_10}
  \end{subfigure}
    \begin{subfigure}[b]{0.45\textwidth}
    \centering
    \includegraphics[width=\textwidth]{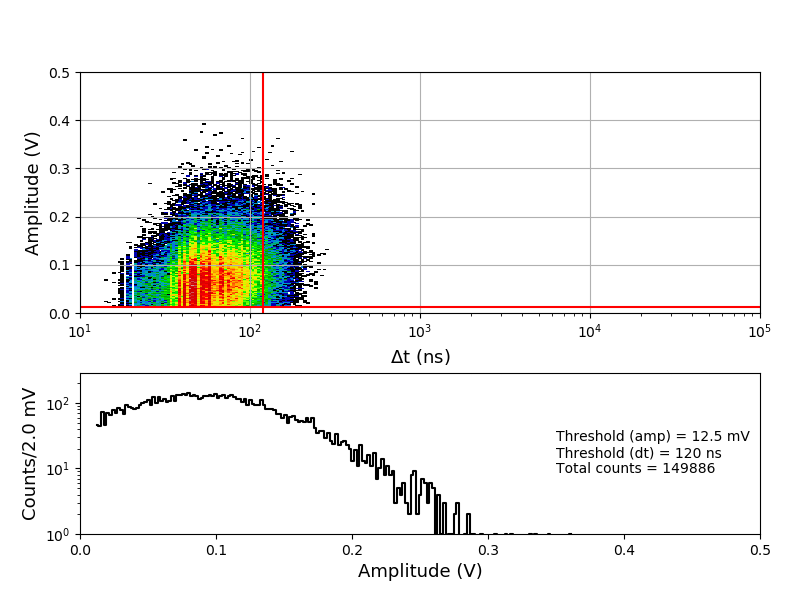}
    \caption{}
    \label{fig:S13_after_m5}
  \end{subfigure}
  \begin{subfigure}[b]{0.45\textwidth}
    \centering
    \includegraphics[width=\textwidth]{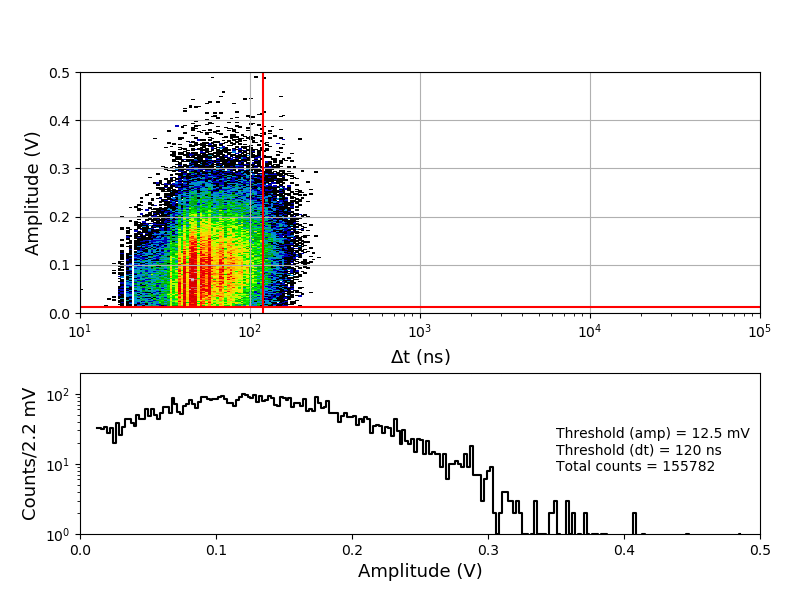}
    \caption{}
    \label{fig:S13_after_m20}
  \end{subfigure}
  \caption{An example of waveform analysis for data taken for S13361 single channel measured at 27$^\circ$C for doses: a) \SI{0.267}{Gy}, b) \SI{0.815}{Gy}, c) \SI{2.19}{Gy} and d) \SI{4.96}{Gy}. The Z-axis is represented here by a colour scale. The overvoltage was set to 2.8~V in each case. }
  \hfill
  \label{fig:S13_DCR_after}
\end{figure}

After the proton irradiation, as in section \ref{sec:dc_before_irradiation}, the \SI{10}{ms} waveforms were also collected at temperatures between $\SI{25}{^{\circ}C}$ and $\SI{-20}{^{\circ}C}$. Figure~\ref{fig:S13_DCR_after} shows the 3D plots and their projections to the y-axis (with the same cuts as before) of a S13361 single channel irradiated for different doses and measured at $\SI{27}{^{\circ}C}$. Due to the significantly higher number of counts the lower overvoltage region was explored. V$_{ov}$ was set to \SI{2.8}{V} in this particular case. As can be seen, it is impossible to extract any single p.e. structure in the DC spectrum. DC rate also increases. \newline 

\begin{figure}[!h]
  \centering
  \begin{subfigure}[b]{0.45\textwidth}
    \centering
    \includegraphics[width=\textwidth]{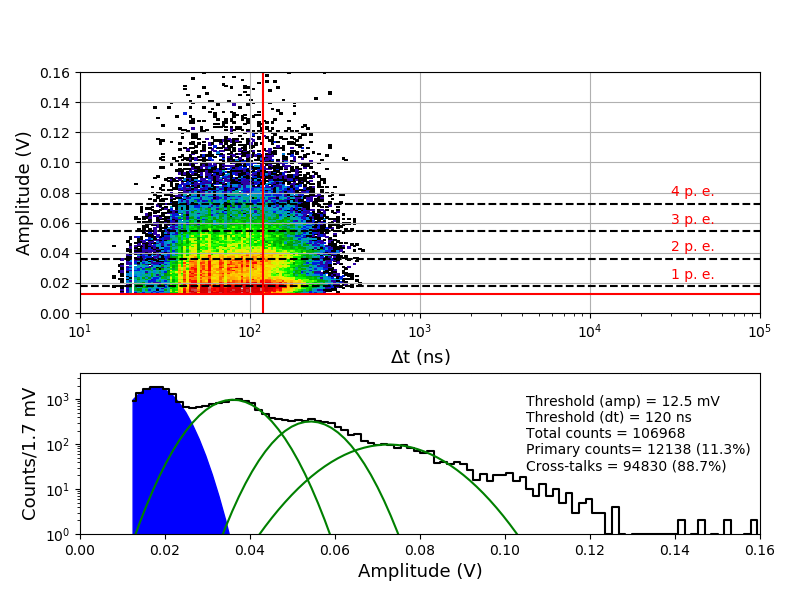}
    \caption{}
    \label{fig:S13_2mths_J4J20_25C_0.4Gy}
  \end{subfigure}
  \begin{subfigure}[b]{0.45\textwidth}
    \centering
    \includegraphics[width=\textwidth]{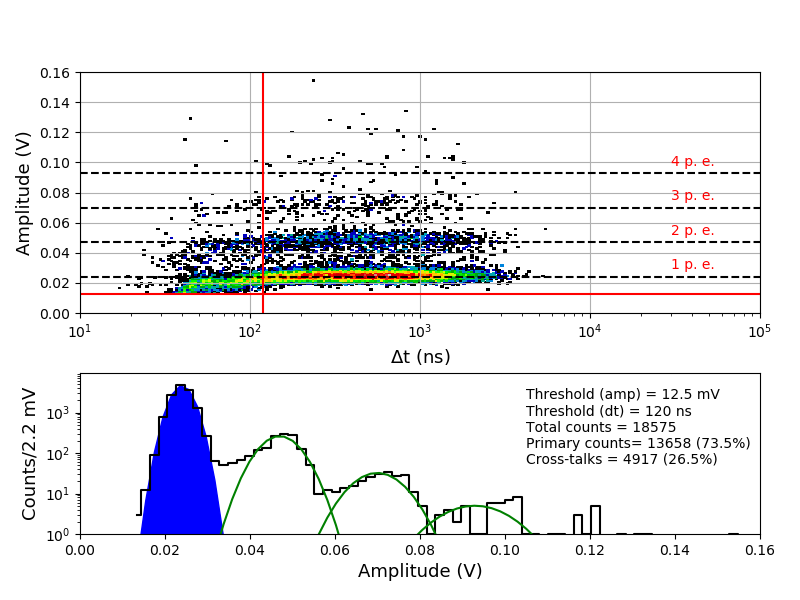}
    \caption{}
    \label{fig:S13_2mths_J9J25_m20C_0.4Gy}
  \end{subfigure}
  \caption{An example of waveform analysis for data taken two months after irradiation with dose \SI{0.267}{Gy} for S13361 single channel at a) $\SI{25}{^{\circ}C}$ and b) $\SI{-20}{^{\circ}C}$. The Z-axis is represented here by a colour scale. The overvoltage was set to \SI{2.8}{V}.}
  \hfill
  \label{fig:S13_DCR_2mths_0.4Gy_temp}
\end{figure}

\begin{figure}[!h]
  \centering
  \begin{subfigure}[b]{0.45\textwidth}
    \centering
    \includegraphics[width=\textwidth]{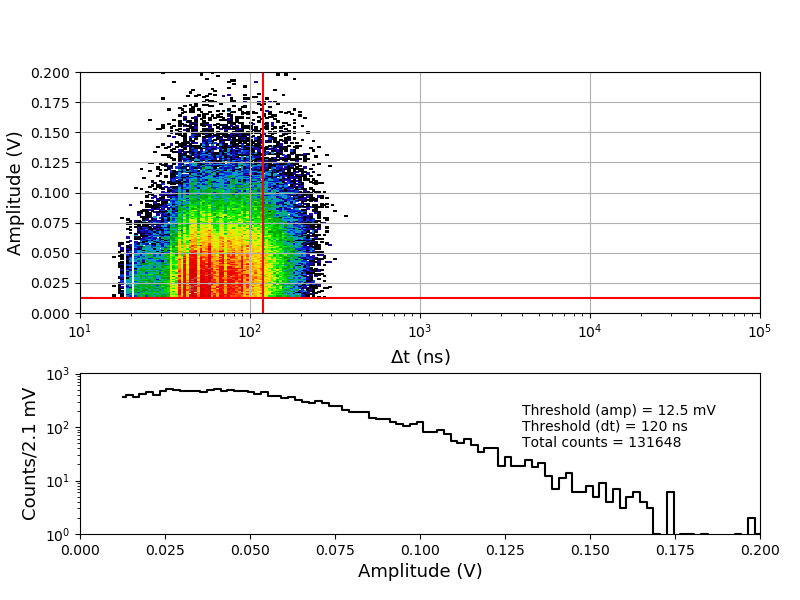}
    \caption{}
    \label{fig:S13_2mths_J4J20_25C}
  \end{subfigure}
  \begin{subfigure}[b]{0.45\textwidth}
    \centering
    \includegraphics[width=\textwidth]{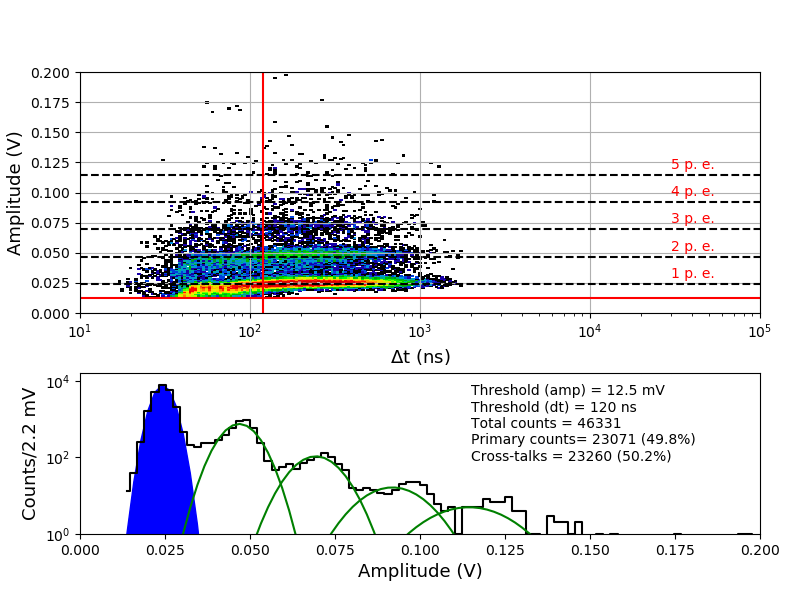}
    \caption{}
    \label{fig:S13_2mths_J9J25_m20C}
  \end{subfigure}
  \caption{An example of waveform analysis for data taken two months after irradiation with dose \SI{0.815}{Gy} for S13361 single channel at a) $\SI{25}{^{\circ}C}$ and b) $\SI{-20}{^{\circ}C}$. The Z-axis is represented here by a colour scale. The overvoltage was set to \SI{2.8}{V}.}
  \hfill
  \label{fig:S13_DCR_2mths_1.2Gy_temp}
\end{figure}

\begin{figure}[!h]
  \centering
  \begin{subfigure}[b]{0.45\textwidth}
    \centering
    \includegraphics[width=\textwidth]{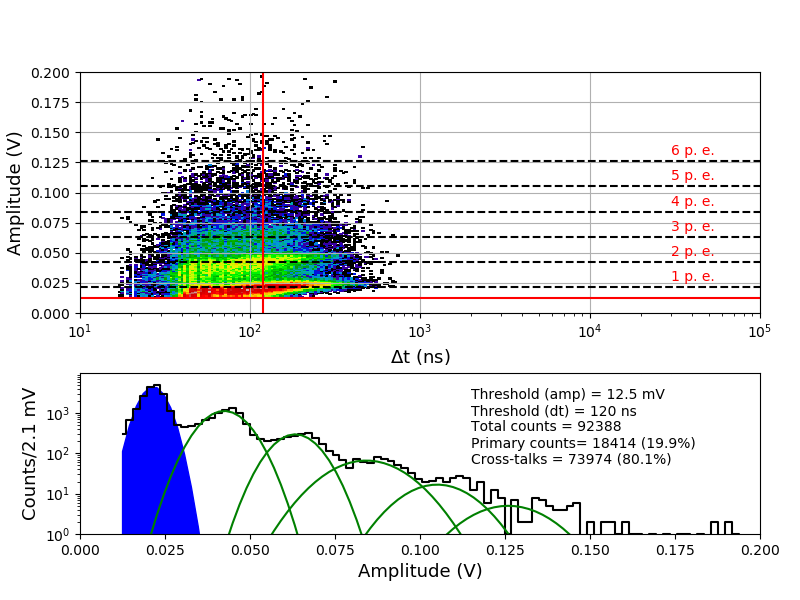}
    \caption{}
    \label{fig:S13_2mths_J4J20_m20C}
  \end{subfigure}
  \begin{subfigure}[b]{0.45\textwidth}
    \centering
    \includegraphics[width=\textwidth]{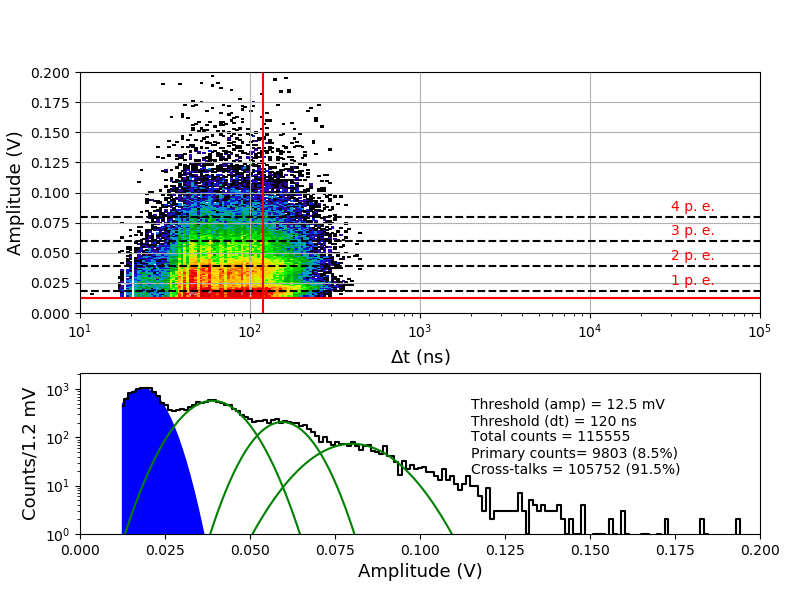}
    \caption{}
    \label{fig:S13_2mths_J16J32_m20C}
  \end{subfigure}
  \caption{An example of waveform analysis for data taken two months after S13361 irradiation with doses a) \SI{2.19}{Gy} and b) \SI{4.96}{Gy} at temperature $\SI{-20}{^{\circ}C}$. The Z-axis is represented here by a colour scale. The overvoltage was set to \SI{2.8}{V}.}
  \hfill
  \label{fig:S13_DCR_2mths_3.5_7.0_Gy_temp}
\end{figure}

The same studies were repeated two months after the proton irradiation (at the same temperatures). For compactness, we only present the data for near opposite temperatures ($\SI{-20}{^{\circ}C}$ vs. $\SI{25}{^{\circ}C}$) at a dose of \SI{0.267}{Gy} (Figure~\ref{fig:S13_DCR_2mths_0.4Gy_temp}) and \SI{0.815}{Gy} (Figure~\ref{fig:S13_DCR_2mths_1.2Gy_temp}). For higher doses of \SI{2.19}{Gy} and \SI{4.96}{Gy} we only present the DC spectra at $\SI{-20}{^{\circ}C}$ (Figure~\ref{fig:S13_DCR_2mths_3.5_7.0_Gy_temp}). It should be noted that when irradiated with a dose of \SI{4.96}{Gy}, contrary to $\SI{25}{^{\circ}C}$, the p.e. peaks are still visible at $\SI{-20}{^{\circ}C}$. \newline

\begin{figure}[!h]
  \centering
  \begin{subfigure}[b]{0.465\textwidth}
    \centering
    \includegraphics[width=\textwidth]{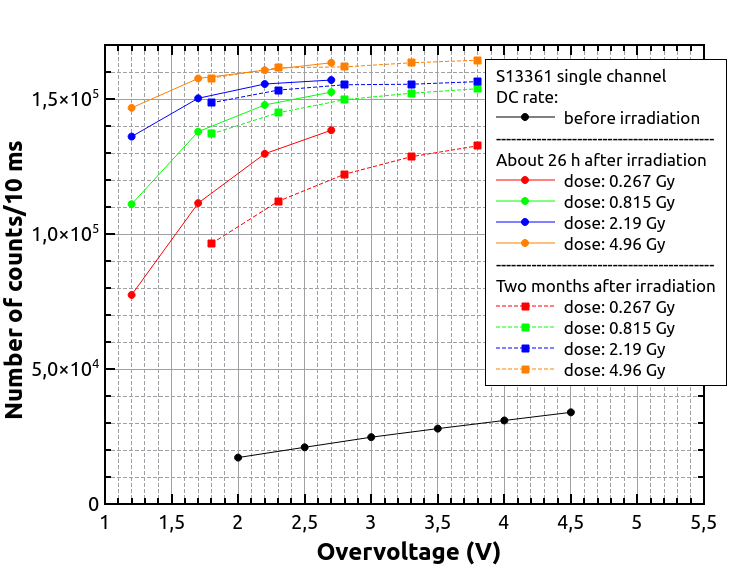}
    \caption{}
    \label{fig:S13_DCR_after_time}
  \end{subfigure}
  \begin{subfigure}[b]{0.45\textwidth}
    \centering
    \includegraphics[width=\textwidth]{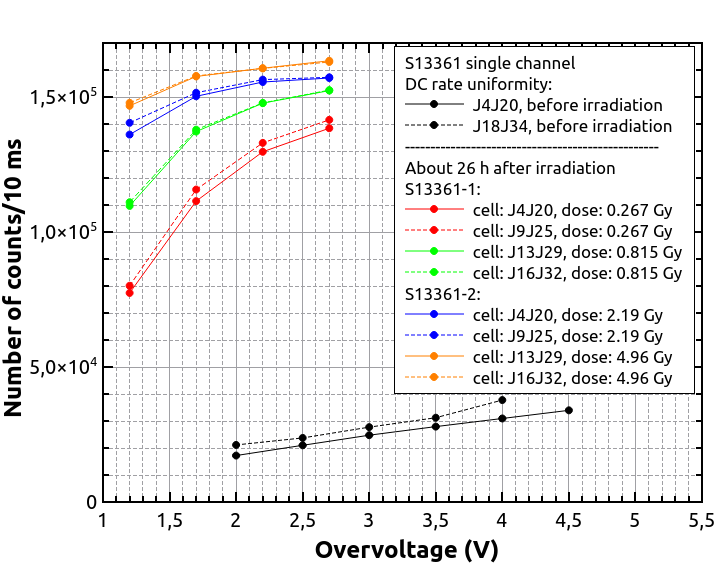}
    \caption{}
    \label{fig:S13_DCR_uniformity}
  \end{subfigure}
  \caption{The number of dark counts per \SI{10}{ms} interval time measured at room temperature for S13361 a) before irradiation, 26~h and two months after b) for two different channels before and \SI{26}{h} after irradiated. Different doses are presented.}
  \hfill
  \label{fig:S13_DCR_general}
\end{figure}

Figure~\ref{fig:S13_DCR_after_time} summarizes the DC rate measured \SI{26}{h} and two months after the proton irradiation for different doses for S13361 at room temperature. 
The uniformity of two different channels (e.g. J4J20 and J9J25 here) of the S13361 is also studied and shown in Figure~\ref{fig:S13_DCR_uniformity}. These channels have been exposed to a dose of \SI{0.267}{Gy} at $\SI{27}{^{\circ}C}$. The number of counts between channel are consistent within a few percent. \newline

\begin{figure}[!h]
  \centering
    \includegraphics[width=\textwidth]{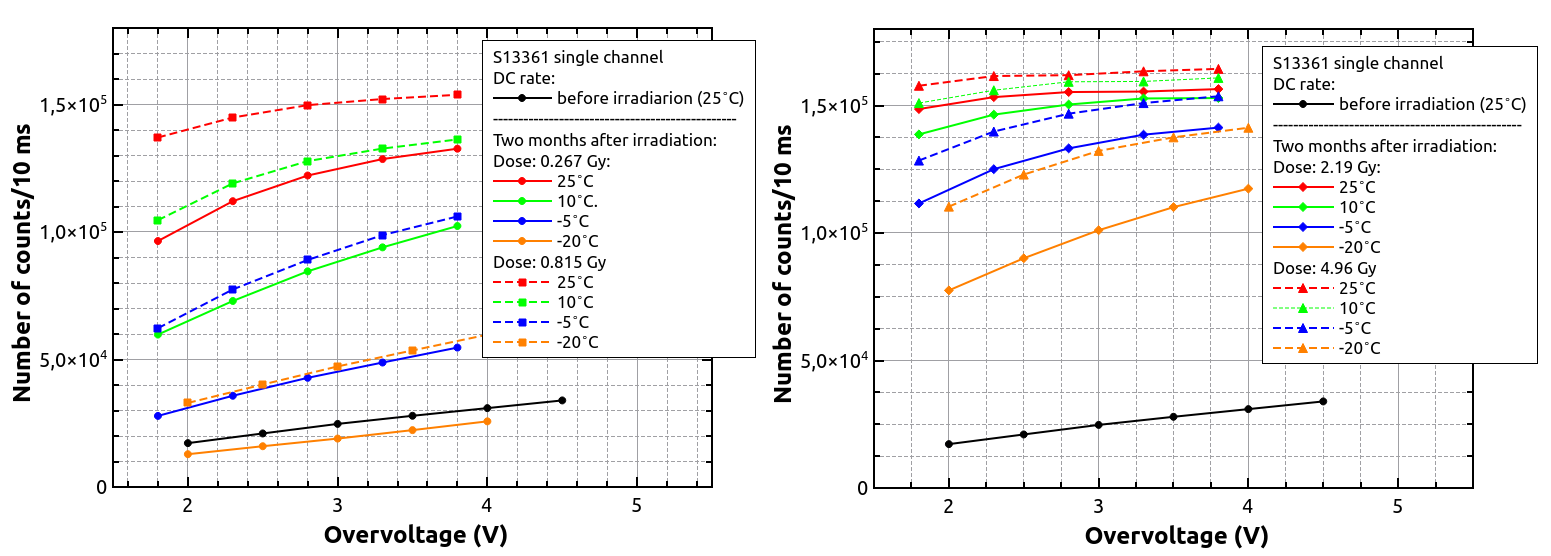}
  \caption{The number of dark counts per \SI{10}{ms} interval time measured before irradiation and two months after for S13361 for different doses and temperatures.}
  \hfill
  \label{fig:S13_DCR_2mths_temp}
\end{figure}

Finally, Figure~\ref{fig:S13_DCR_2mths_temp} shows the number of dark counts measured for different doses at different temperatures. In each case the decreasing number of counts with respect to temperature decrease is clearly visible. The direct correlation between DC and temperature is expected and stresses the importance of temperature control of SiPMs for POLAR-2, thus (also) minimizing the impact of radiation damage. \newline

\begin{figure}[!h]
  \centering
  \begin{subfigure}[b]{0.53\textwidth}
    \centering
    \includegraphics[width=\textwidth]{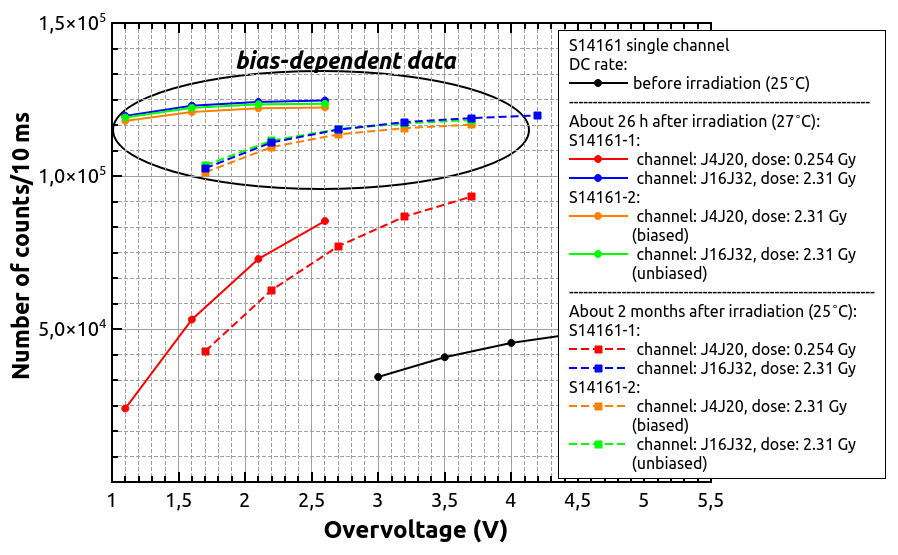}
    \caption{}
    \label{fig:S14_DCR_after}
  \end{subfigure}
  \begin{subfigure}[b]{0.45\textwidth}
    \centering
    \includegraphics[width=\textwidth]{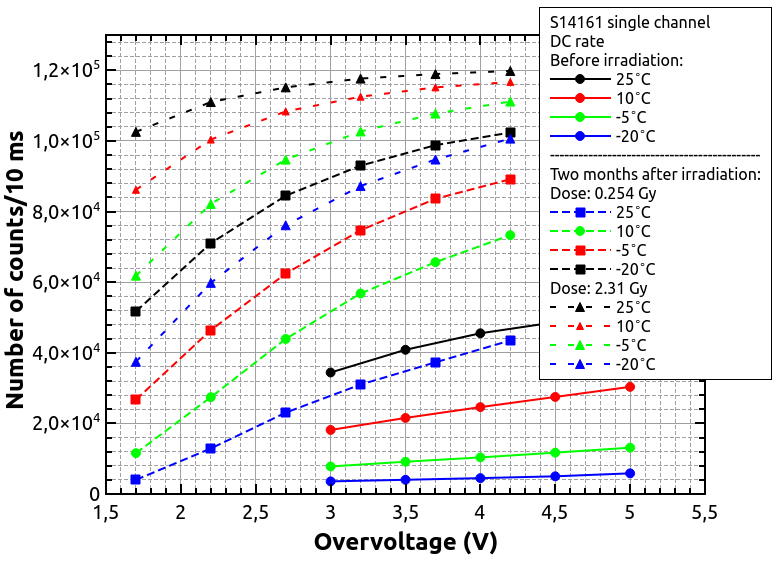}
    \caption{}
    \label{fig:S14_DCR_2mths}
  \end{subfigure}
  \caption{The number of dark counts per \SI{10}{ms} interval time measured for S14161 a) before irradiation, 26~h and 2~months after at room temperature b) for two different channels irradiated with the same dose 26~h after proton irradiation. As it is shown in a) the bias-dependent effect is not visible.}
  \hfill
  \label{fig:S14_DCR_general}
\end{figure}

Similarly to Figure~\ref{fig:S13_DCR_2mths_temp}, Figure~\ref{fig:S14_DCR_general} summarizes the obtained data for the S14161 array. The selected irradiation scenarios are those with dose of 0.254 and \SI{2.31}{Gy}. Again, the figure shows the number of counts determined for a \SI{10}{ms} interval time. In general, the data signifies a good SiPM single channel uniformity. Also, considering the I-V characteristics, no bias-dependent effect during proton irradiation was observed. Finally, the number of counts is also self-consistent between different single channels (exposed to the same dose). \newline

\begin{figure}[!h]
  \centering
  \includegraphics[width=0.6\textwidth]{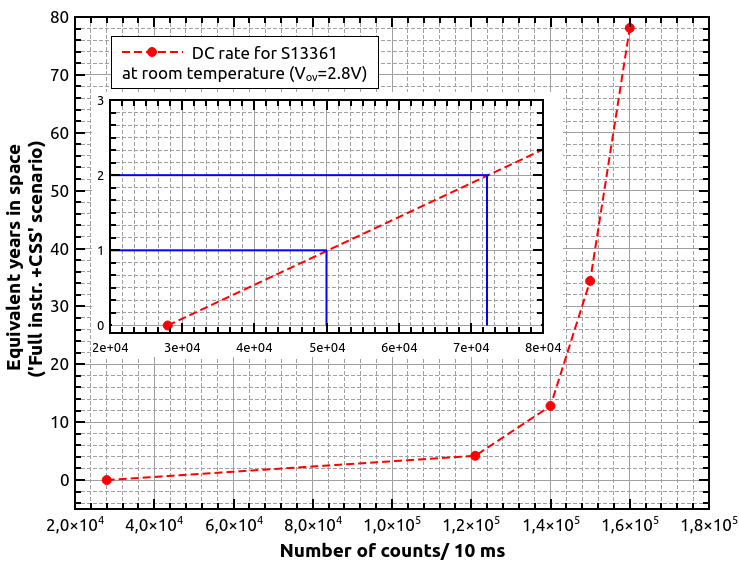}
  \caption{The number of DC measured at room temperature and V$_{ov}$=\SI{2.8}{V}, when 'Full Instrument + CSS' scenario is assumed (more details in the text). The inset shows expected DC rate after one and two years equivalent in cosmic space.} 
  \label{fig:DCR_interp}
\end{figure}

Figure~\ref{fig:DCR_interp} shows our estimation of DC rates (based on a simple interpolation) for $\sim$1 and $\sim$2 years in space, assuming the 'Full Instrument + CSS' scenario at room temperature (V$_{ov}$=\SI{2.8}{V}). The presented numbers correspond to \SI{26}{h} after proton irradiation. At the stage, the annealing effect and consequently the DC decreasing are negligible. We therefore conclude, that numbers of about \SI{5.0e6}{counts/second} (cps) and \SI{7.2e6}{cps} (one and two years equivalent, respectively) should not influence the SiPM single channel operation. Nevertheless, at the same conditions, the maximum dark pulse amplitudes increased from about \SI{110}{mV} for the cases before irradiation, to about \SI{150}{mV} when irradiated with a dose of \SI{0.267}{Gy}. This implies that, for POLAR-2, the threshold must be carefully selected. \newline

\newpage

\subsection{Activation analysis} \label{sec:activation}

The high energy resolution of High Purity Germanium detectors (HPGe) provide a way to identify characteristic gamma rays from proton activation products within the SiPM(+PCB) array. In the case of S13361, immediately after it was irradiated with a dose of \SI{4.96}{Gy} , the SiPM array was moved to the HPGe detector to measure the most prominent gamma lines. The HPGe was placed inside a low-background lead protected chamber to decrease background radiation. \newline

The signal readout from the HPGe detector was based on analogue NIM electronics, where the signal from the HPGe preamplifier was shaped and amplified by an ORTEC spectroscopy amplifier. Finally, the signal was registered by a multichannel analyzer which writes the energy spectrum to file in \SI{100}{s} intervals. \newline

\SI{10}{min} was required to move the irradiated sample from the experimental hall and position it into the HPGe setup, introducing a non-negligible time delay between the irradiation process and the start of data acquisition. This limits our detection ability of decay products with decay times shorter than 2-3 minutes.\newline

Many of the radioisotopes in reaction with \SI{58}{MeV} protons have multiple modes of decay, where daughter nuclei mostly decay into $\beta^+$. This explains the strong population of counts at the \SI{511}{keV} line. To limit the number of decay modes, the isotopes are identified through a full-energy peak fitting of the measured HPGe spectrum. This facilitated the determination of the centroid/energy of the gamma transition inside the isotope. 

The TALYS \cite{talys} software was used to simulate the nuclear reactions from \SI{58}{MeV} protons. By knowing the energy of incoming protons and the main components of the SiPM array and PCB plate, such as: carbon, silicon, oxygen, potassium, etc., the tables of possible decay channels with total cross-sections were generated. Based on the highest cross-section criterion, we listed in Table~\ref{tab:activ} the most prominent decay modes and corresponding decay times \cite{NuDAT}. The absence of silicon and oxygen in the Table~\ref{tab:activ} may be surprising. However, for silicon, the decay time is in the range of a few seconds. For oxygen, the most prominent decay channel has a decay time of \SI{122}{s}. Its gamma-ray contribution is only to the \SI{511}{keV} peak. As a result, to a first order, these components are neglected. \newline 

\begin{table}[ht]
\centering
\begin{tabular}{ | c | c |c |c |c |c |c | }
\hline
Mother & Abundance  & Daughter & Reaction & Total cross & T$_{1/2}$  & Dominant      \\ 
nuclei & ($\%$)    & nuclei   & type     & section (mb) & (s)		  & energies (keV) \\
\hline
Sn-120 & 32.6 & Sb-115 & (p,6n) & 223.6 & 1926 & 511 \\ 
\hline
Cu-63 & 69.0 & Cu-60 & (p,p3n) & 27.8 & 1422 & 511, 826, 1333, 1792 \\ 
\hline
Cu-63 & 69.0 & Cu-61 & (p,p2n) & 92.8 & 12010 & 511 \\ 
\hline
Cu-63 & 69.0 & Cu-62 & (p,pn) & 167 & 580 & 511 \\ 
\hline
Cu-65 & 31.0 & Cu-64 & (p,pn) & 167 & 45724 & 511 \\ 
\hline
Cu-65 & 31.0 & Cu-62 & (p,p3n) & 117 & 580 & 511 \\ 
\hline
C-12 & 99.0 & C-11 & (p,pn) & 167 & 1222 & 511 \\ 
\hline
K-39 & 93.0 & K-38 & (p,pn) & 67.3 & 459 & 511, 2168 \\ 
\hline
\end{tabular}
\caption{Dominant reactions of \SI{58}{MeV} protons on SiPM(+PCB) based elements calculated using TALYS.}
\label{tab:activ}
\end{table}

\begin{figure}[!h]
  \centering
  \begin{subfigure}[b]{0.48\textwidth}
    \centering
    \includegraphics[width=\textwidth]{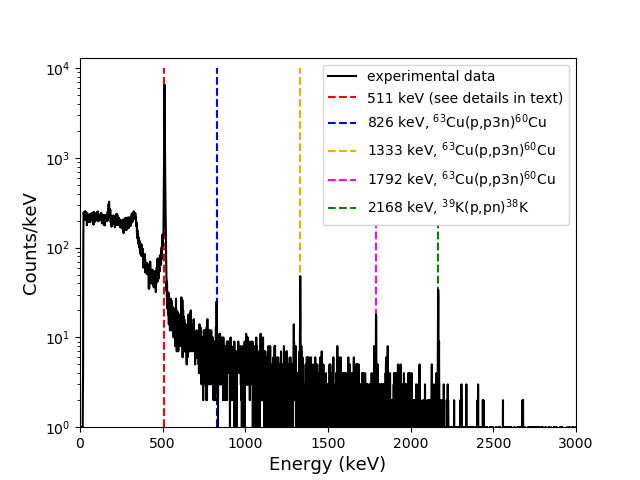}
    \caption{}
    \label{fig:energy_spec}
  \end{subfigure}
  \begin{subfigure}[b]{0.48\textwidth}
    \centering
    \includegraphics[width=\textwidth]{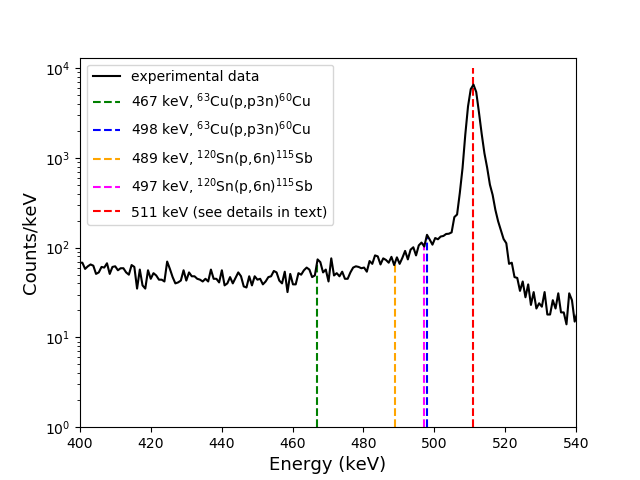}
    \caption{}
    \label{fig:energy_spec_low}
  \end{subfigure}
  \caption{The example of gamma-ray energy spectrum measured with HPGe detector after SiPM proton irradiation. Black line shows experimental data. Dashed colour lines show identified peaks.}
  \hfill
  \label{fig:e_spec_activ}
\end{figure}

Figure~\ref{fig:e_spec_activ} shows a typical gamma-ray energy spectrum from the HPGe detector. The spectrum shown in Figure~\ref{fig:energy_spec} accounts for the full energy range, whereas Figure~\ref{fig:energy_spec_low} highlights the spectral structure near \SI{511}{keV}. We observe five strong gamma lines, corresponding primarily to copper and potassium activation products. It can be also seen that to the left of the \SI{511}{keV} line, some additional small contributions are present. The TALYS calculations show that the sources of these lines are copper and zinc but emitted with much lower intensities. We would like to note that background contribution is negligible in the presented spectrum. Our long-time background measurement showed, that the highest contribution in energy spectrum for energies starting from \SI{400}{keV}, was below 0.2 counts per \SI{100}{s}. \newline

\begin{figure}[!h]
  \centering
  \begin{subfigure}[b]{0.45\textwidth}
    \centering
    \includegraphics[width=\textwidth]{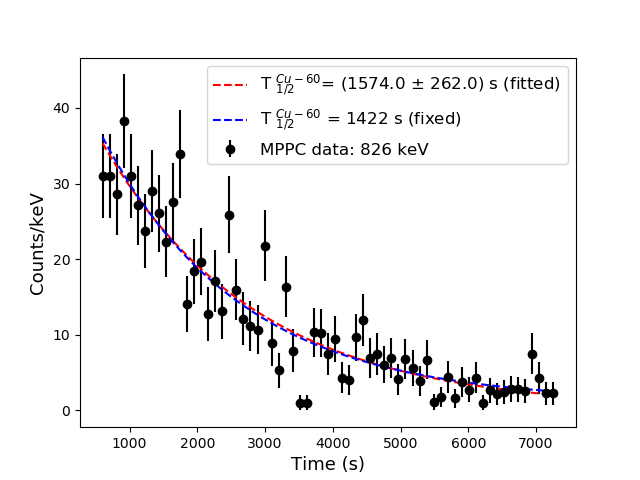}
    \caption{}
    \label{fig:decay_time_826keV}
  \end{subfigure}
  \begin{subfigure}[b]{0.45\textwidth}
    \centering
    \includegraphics[width=\textwidth]{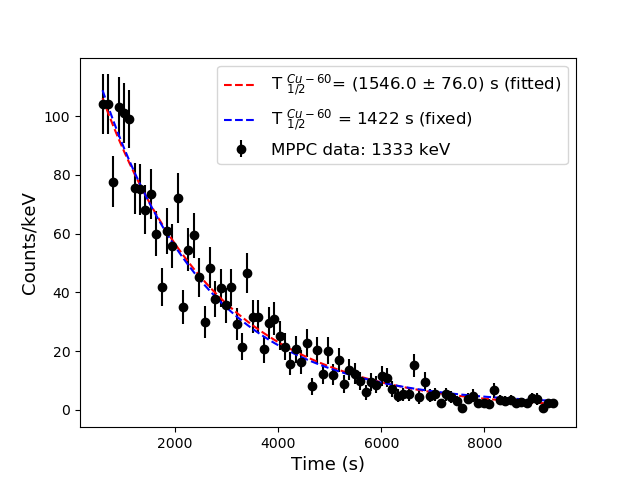}
    \caption{}
    \label{fig:decay_time_1333keV}
  \end{subfigure}
    \begin{subfigure}[b]{0.45\textwidth}
    \centering
    \includegraphics[width=\textwidth]{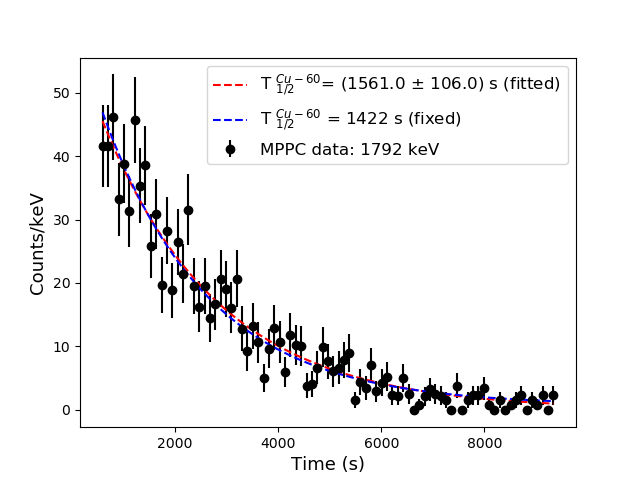}
    \caption{}
    \label{fig:decay_time_1792keV}
  \end{subfigure}
  \begin{subfigure}[b]{0.45\textwidth}
    \centering
    \includegraphics[width=\textwidth]{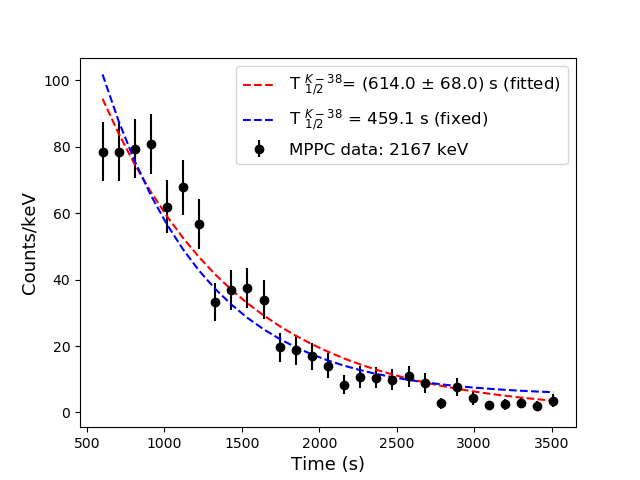}
    \caption{}
    \label{fig:decay_time_2167keV}
  \end{subfigure}
  \caption{Decay time distributions (black points) measured for \SI{58}{MeV} proton irradiated SiPM. Subfigures corresponds to different energy gated lines a) \SI{826}{keV}, b) \SI{1333}{keV}, c) \SI{1792}{keV} and d) \SI{2167}{keV}. Red dashed lines correspond to fitting procedure, where amplitudes and decay times were set as a free parameters. Blue dashed lines describe the cases with fixed decay times (see details in text).}
  \hfill
  \label{fig:decay_time}
\end{figure}

To confirm our peak's identification the decay times for certain energies were determined experimentally. Figure~\ref{fig:decay_time} shows the obtained results. Due to limited count statistics, the obtained uncertainties are relatively high; over 50$\%$ in some cases. For this reason we decided to focus on two scenarios. In the first one, the decay time and amplitude were free fitting parameters (red dashed line). In the second scenario, a certain fixed decay time was taken from the Table~\ref{tab:activ} (blue dashed line) and only the amplitude was fitted. Generally, fitted curves in the second scenario are in good agreement with the experimental data. The obtained decay times are also in good agreement (within 3$\sigma$) for the lines produced by Sn-120, CU-63, Cu-65, C-12 and K-39, confirming our identification. \newline

\begin{figure}[!h]
  \centering
  \begin{subfigure}[b]{0.45\textwidth}
    \centering
    \includegraphics[width=\textwidth]{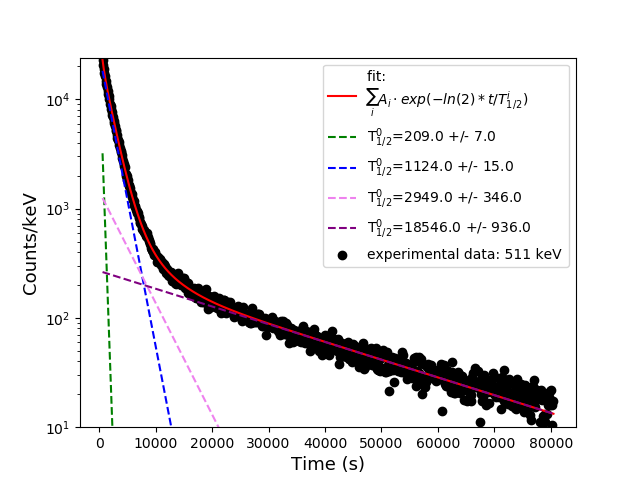}
    \caption{}
    \label{fig:decay_511keV_free}
  \end{subfigure}
  \begin{subfigure}[b]{0.45\textwidth}
    \centering
 	 \includegraphics[width=\textwidth]{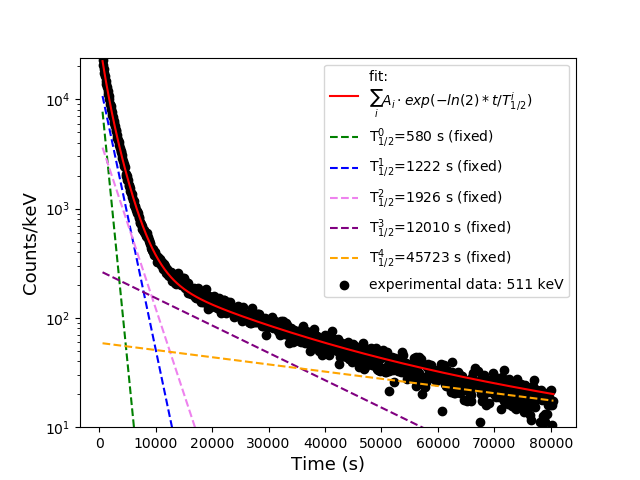}
 	 \caption{}
 	 \label{fig:decay_511keV_fixed}
  \end{subfigure}
  \caption{Decay time distribution of \SI{511}{keV} line (black points) measured for \SI{58}{MeV} proton irradiated SiPM. Red lines show the results of fitting procedure for a) N=4  (8 free parameters) and b) N=5 (5 free parameters - only amplitudes). Dashed lines show each component contribution (see more details in text).}
  \hfill
  \label{fig:decay_511keV}
\end{figure}

As is shown in Table~\ref{tab:activ}, there are many potential sources for the \SI{511}{keV} line. This is reflected in the shape of the decay time distribution in Figure~\ref{fig:decay_511keV} (black points), where at least two (fast and long tail) components are clearly visible.
To describe the experimental data the sum of N exponential components were tested:

\begin{equation} \label{eq:decay}
y = \sum_i^N A_i \cdot exp\left(-\frac{t\cdot ln(2)}{T_{1/2}^i}\right),
\end{equation}

 where $A_{i}$ and $T_{1/2}^i$ refer to the amplitude and the decay time of the component respectively. To obtain a good description of the experimental data, 4 components (N=4) were used, resulting in 8 free parameters. The decay times obtained during the fitting procedure, presented in Figure~\ref{fig:decay_511keV_free}, are relatively far away from the expected values presented in Table~\ref{tab:activ} (over 3$\sigma$) for the most dominant nuclei and does not match to other potential candidates from other nuclei. This could be attributed to low statistics in the tail component, biasing towards shorter time bases, or the incorrect estimation of the number of element contributions in the \SI{511}{keV} peak. The simplest approach N$>$4 increased the number of free parameters and significantly complicates the fitting procedure causing even higher uncertainties (overfitting and/or correlated parameters). \newline

In the second approach, it was decided to use the decay times listed in Table \ref{tab:activ} as fixed parameters while only varying the amplitudes. Cu-63 and Cu-65 have the same daughter nuclei, Cu-62, reducing the number of components in Equation~\ref{eq:decay} by one. Next, it was assumed that the contributions of Cu-60 and K-39, which have the smallest total cross section from those presented in Table~\ref{tab:activ}, are also negligible, despite seeing other gamma lines from this channel. This means that for 5 components (N=5) we only fit 5 free parameters (as opposed to 10). The result of this fitting procedure is shown in Figure~\ref{fig:decay_511keV_fixed}. It can be seen, that based on TALYS calculations, we are in good agreement with experimental data even with a smaller number of free parameters. This allows us to conclude, that our assumption about basic SiPM(+PCB) element composition is correct and we can accurately modeling the setup. \newline

POLAR-2 will orbit with a period of \SI{90}{min}, frequently passing through the SAA\footnote{Due to the orbital configuration it will pass the SAA at different regions.}. From the figures \ref{fig:decay_511keV_free} and \ref{fig:decay_511keV_fixed} we see that most elements have already decayed, reducing the count number by at least one order of magnitude. As the SiPM (used to derive the activation studies) was irradiated with \SI{4.96}{Gy}, equating to \SI{1.78}{years} of equivalent space time for the 'bare' scenario (see Table~\ref{tab:rad_dose}), we see that the number of counts already drop by more than three orders of magnitude after a day. For POLAR-2, the dose equates to \SI{62.9}{years} of equivalent time in space. Furthermore, we see that the most dominating components (irrespective of fitting procedure) have decay times shorter than the orbital period. As a result, we do not expect a significant contribution of the activation of the SiPMs and the PCB to the dose to which the sensors are exposed. \newline

\newpage
\section{Summary and outlook}

We have presented here the results of the irradiation of the S13361-6075NE-04 and S14161-6050HS-04 SiPM arrays from Hamamatsu. The campaign was performed for the POLAR-2 mission which is planned for launch on the CSS in late 2024 or early 2025. Two scenarios were studied: i) the 'Full Instrument + CSS' for POLAR-2 purposes and ii) the 'Bare SiPM' scenario for other instruments which may be more exposed to the background radiation. \newline

After an exposure of \SI{4.96}{Gy}, it was found that the S13361 SiPM array is still operational. For the 'Full Instrument + CSS' scenario, this equates to \SI{62.9}{years} in space, whereas it equates to \SI{1.78}{years} for the 'Bare SiPM' scenario. The most prominent effects after proton irradiation are the increase in dark current and dark counts, mostly by cross-talk events. The observed changes in the dark count rate, compared to data taken before irradiation, are about 4 times higher with respect to the lowest dose. Furthermore, the maximum dark pulse amplitudes for V$_{ov}$=\SI{2.8}{V} increased from about \SI{110}{mV} (the last p.e. peak at $\SI{25}{^{\circ}C}$, before irradiation) to about \SI{350}{mV} (left edge of DC spectrum, $\SI{27}{^{\circ}C}$, \SI{4.96}{Gy}), showing the necessity of a well-thought threshold determination. These effects can be reduced significantly by lowering the SiPM temperature. The maximum dark pulse amplitude for the dose \SI{4.96}{Gy} and the temperature  $\SI{-20}{^{\circ}C}$ is about \SI{150}{mV}, that is close to a value before the irradiation (at room temperature). \newline

Annealing processes were observed for DC spectra taken after \SI{2}{hours}, \SI{26}{hours} and \SI{2}{months}. Although not the emphasis of this paper, they are later specified through a dedicated study and will be published hereafter. 
Neither SiPM type showed a change in performance when a single channel was biased (or unbiased) during proton irradiation as the measured I-V characteristics and DC rates did not expose significant differences in this case. Nevertheless, it is recommended, despite not affecting the radiation damage, to reduce the bias voltage in the SAA region to avoid issues with the SiPM current as well as the produced data volume. \newline

Finally, we discussed the SiPM(+PCB) proton activation analysis. The results show a dominant contribution of the \SI{511}{keV} line in the energy spectrum measured with the HPGe detector, corresponding to $\beta^+$ decay of daughter nuclei populated in proton reaction. Excluding life times in the range of seconds (due to practical limitations), the main contribution in the gamma spectrum comes from copper and carbon activation which mostly have a decay time shorter than the orbital period. As a result, we do not anticipate the activation products from SiPMs to provide a significant contribution in the polarimeter response. 

In conclusion, due to the shielding in the 'Full Instrument + CSS' scenario, we do not expect POLAR-2 to greatly suffer from a rapidly degrading SiPM through background radiation. \newline

\newpage

\bmhead{Acknowledgments}

We gratefully acknowledge the Swiss Space Office of the State Secretariat for Education, Research and Innovation (ESA PRODEX Programme) which supported the development and production of the POLAR-2 detector. M.K. and N.D.A. acknowledge the support of the Swiss National Science Foundation. National Centre for Nuclear Research acknowledges support from Polish National Science Center under the grant UMO-2018/30/M/ST9/00757. We gratefully acknowledge the support from the National Natural Science Foundation of China (Grant No. 11961141013, 11503028), the Xie Jialin Foundation of the Institute of High Energy Physics, Chinese Academy of Sciences (Grant No. 2019IHEPZZBS111), the Joint Research Fund in Astronomy under the cooperative agreement between the National Natural Science Foundation of China and the Chinese Academy of Sciences (Grant No. U1631242), the National Basic Research Program (973 Program) of China (Grant No. 2014CB845800), the Strategic Priority Research Program of the Chinese Academy of Sciences (Grant No. XDB23040400), and the Youth Innovation Promotion Association of Chinese Academy of Sciences. We would like to thank T.~Horwacik, T.~Nowak and L.~Malinowski for their support during the irradiation campaign at IFJ.

\newpage
\section*{Declarations}

\subsection*{Funding}

The (co-)authors are funded by the funding the agencies described in the acknowledgment section. 

\subsection*{Conflicts of Interest}
The authors declare that the research was conducted in the absence of any commercial or financial relationships that could be construed as a potential conflict of interest.

\subsection*{Consent to participate} 
Not applicable.

\subsection*{Consent for publication} 
Not applicable.

\subsection*{Code availability}
Not applicable.

\subsection*{Author's Contribution}
The main three authors are Slawomir Mianowski, Nicolas De Angelis and Johannes Hulsman. Slawomir Mianowski is considered the primary of the three. The first draft of the manuscript was written by Slawomir Mianowski, Nicolas De Angelis and Johannes Hulsman and all authors commented on previous versions of the manuscript.


\newpage
\bibliography{sn-bibliography}


\end{document}